\documentclass[12pt]{article}
\usepackage{epsfig}
\usepackage{axodraw}
\usepackage{a4}
\usepackage{latexsym}

\usepackage{pslatex}
\usepackage[latin1]{inputenc}
\usepackage[T1]{fontenc}

\usepackage{color,colordvi}
\def\colour4colour#1{\Blue{#1}}

\newcommand{\lsim}{\raisebox{-0.07cm}{$\:\:\stackrel{<}{{\scriptstyle
 \sim}}\:\: $} }
\newcommand{\beq}{\begin{equation}}
\newcommand{\eeq}{\end{equation}}
\newcommand{\bea}{\begin{eqnarray}}
\newcommand{\eea}{\end{eqnarray}}
\newcommand{\nn}{\nonumber}
\newcommand{\MSb}{$\overline{\mbox{MS}}$}
\newcommand{\as}{\alpha_{\rm s}}
\newcommand{\ra}{\rightarrow}
\newcommand{\DD}{{\cal D}}

\begin{document}
\setlength{\parskip}{0.3cm}
\setlength{\baselineskip}{0.55cm}

\def\plus{{\!+\!}}
\def\minus{{\!-\!}}
\def\z#1{{\zeta_{#1}}}
\def\ca{{C^{}_A}}
\def\cf{{C^{}_F}}
\def\nf{{n^{}_{\! f}}}
\def\n2f{{n^{\,2}_{\! f}}}

\def\dabc2n{{{d^{abc}d_{abc}}\over{n_c}}}
\def\S(#1){{{S}_{#1}}}
\def\Ss(#1,#2){{{S}_{#1,#2}}}
\def\Sss(#1,#2,#3){{{S}_{#1,#2,#3}}}
\def\Ssss(#1,#2,#3,#4){{{S}_{#1,#2,#3,#4}}}
\def\Sssss(#1,#2,#3,#4,#5){{{S}_{#1,#2,#3,#4,#5}}}
\def\Npm{{{\bf N_{\pm}}}}
\def\Npmi{{{\bf N_{\pm i}}}}
\def\Nminus{{{\bf N_{-}}}}
\def\Nplus{{{\bf N_{+}}}}
\def\Nminustwo{{{\bf N_{-2}}}}
\def\Nplustwo{{{\bf N_{+2}}}}
\def\Nminusthree{{{\bf N_{-3}}}}
\def\Nplusthree{{{\bf N_{+3}}}}

\def\pqq(#1){p_{\rm{qq}}(#1)}
\def\H(#1){{\rm{H}}_{#1}}
\def\Hh(#1,#2){{\rm{H}}_{#1,#2}}
\def\Hhh(#1,#2,#3){{\rm{H}}_{#1,#2,#3}}
\def\Hhhh(#1,#2,#3,#4){{\rm{H}}_{#1,#2,#3,#4}}

\begin{titlepage}
\noindent
DESY 04--047 \hfill {\tt hep-ph/0403192}\\
SFB/CPP-04-09 \\
NIKHEF 04-001 \\
March 2004 \\
\vspace{1.3cm}
\begin{center}
\Large
{\bf The Three-Loop Splitting Functions in QCD:} \\
\vspace{0.15cm}
{\bf The Non-Singlet Case} \\
\vspace{1.5cm}
\large
S. Moch$^{\, a}$, J.A.M. Vermaseren$^{\, b}$ and A. Vogt$^{\, b}$\\
\vspace{1.2cm}
\normalsize
{\it $^a$Deutsches Elektronensynchrotron DESY \\
\vspace{0.1cm}
Platanenallee 6, D--15735 Zeuthen, Germany}\\
\vspace{0.5cm}
{\it $^b$NIKHEF Theory Group \\
\vspace{0.1cm}
Kruislaan 409, 1098 SJ Amsterdam, The Netherlands} \\
\vspace{3.0cm}
\large
{\bf Abstract}
\vspace{-0.2cm}
\end{center}
We compute the next-to-next-to-leading order (NNLO) contributions to 
the three splitting functions governing the evolution of unpolarized 
non-singlet combinations of quark densities in perturbative QCD. 
Our results agree with all partial results available in the literature. 
We find that the correct leading logarithmic (LL) predictions for 
small momentum fractions $x$ do not provide a good estimate of the 
respective complete results. A new, unpredicted LL contribution is 
found for the colour factor $d^{abc\,}d_{abc}$ entering at 
three loops for the first time. 
We investigate the size of the corrections and the stability of the
NNLO evolution under variation of the renormalization scale. Except for
very small $x$ the corrections are found to be rather small even for 
large values of the strong coupling constant, in principle facilitating 
a perturbative evolution into the sub-GeV regime.
\vfill
\end{titlepage}
%
%
\section{Introduction}
\label{sec:introduction}
%
%
Parton distributions form indispensable ingredients for the analysis 
of all hard-scattering processes involving initial-state hadrons. The 
dependence of these quantities on the fraction $x$ of the hadron 
momentum carried by the quark or gluon cannot be calculated in 
perturbation theory. However, the scale-dependence (evolution) of the 
parton distributions can be derived from first principles in terms of 
an expansion in powers of the strong coupling constant $\as$. The 
corresponding $n\,$th-order coefficients governing the evolution 
are referred to as the $n$-loop anomalous dimensions or splitting 
functions.  Parton densities evolved by including the terms up to order 
$\as^{\, n+1}$ in this expansion constitute, together with the 
corresponding results for the partonic cross sections for the 
observable under consideration, the N$^{\rm n}$LO (leading-order, 
next-to-leading-order, next-to-next-to-leading-order, etc.) 
approximation of perturbative QCD.

Presently the next-to-leading-order is the standard approximation for
most important processes. The corresponding one- and two-loop splitting
functions have been known for a long time 
\cite{Gross:1973rr,Georgi:1974sr,Altarelli:1977zs,Floratos:1977au,%
Floratos:1979ny,Gonzalez-Arroyo:1979df,Gonzalez-Arroyo:1980he,%
Curci:1980uw,Furmanski:1980cm,Floratos:1981hs,Hamberg:1992qt}.
The NNLO corrections need to be included, however, in order to arrive at
quantitatively reliable predictions for hard processes at present and
future high-energy colliders. These corrections are so far known only
for structure functions in deep-inelastic scattering 
\cite{ vanNeerven:1991nn, Zijlstra:1991qc,Zijlstra:1992kj,%
Zijlstra:1992qd}
and for Drell-Yan lepton-pair and gauge-boson production in 
proton--(anti-)proton collisions 
\cite{Hamberg:1991np,Harlander:2002wh,Anastasiou:2003yy,%
Anastasiou:2003ds}
and the related cross sections for Higgs production in the 
heavy-top-quark approximation 
\cite{Harlander:2002wh,Anastasiou:2002yz,Ravindran:2003um,%
Harlander:2003ai}.
Work on NNLO cross sections for jet production is under way and 
expected to yield results in the near future, see 
Ref.~\cite{Glover:2002gz} and references therein.
For the corresponding three-loop splitting functions, on the other hand, 
only partial results have been obtained up to now, most notably the 
lowest six/seven (even or odd) integer-$N$ Mellin moments 
\cite{Larin:1994vu,Larin:1997wd,Retey:2000nq}.

These Mellin moments already provide a rather accurate description of 
the splitting functions at large momentum fractions $x$ 
\cite{Larin:1997wd,vanNeerven:1999ca,vanNeerven:2000uj,%
vanNeerven:2000wp}. Their much-debated behaviour at small values of 
$x$, on the other hand, can only be determined by a full calculation. 
As we will demonstrate below for the non-singlet cases, this statement 
holds despite the existence of resummation predictions for the leading 
small-$x$ logarithms \cite{Kirschner:1983di,Blumlein:1996jp}, since
{\bf --a--} 
the correctly predicted logarithms do not dominate the three-loop 
splitting functions at any practically relevant value of $x$ and 
{\bf --b--} 
a term of the same size occurs with a new colour factor at third order 
which could not have been predicted from lower-order results,
analogous to the situation for the four-loop $\beta$-function of 
QCD~\cite{vanRitbergen:1997va}.

In this article we present the (unpolarized) flavour non-singlet (ns) 
splitting functions at the third order in perturbative QCD. The 
corresponding flavour singlet results will appear in a forthcoming 
publication~\cite{MVV4}. 
The present article is organized as follows: In section 2 we set up our 
notations for the three independent third-order splitting functions and 
briefly discuss the method of our calculation. The Mellin-$N$ space 
results are written down in section 3 together with their explicit 
large-$N$ limit which is relevant for the soft-gluon threshold 
resummation \cite{Sterman:1987aj,Catani:1989ne,Catani:1991rp} at 
next-to-next-to leading logarithmic accuracy~\cite{Vogt:2000ci}.
A surprising relation is found between the leading large-$N$ term at
two loops and the subleading $(\ln N)/N$ contribution at third order.
In section 4 we present the exact results as well as compact 
parametrizations for the $x$-space splitting functions and study their
behaviour at small~$x$. The numerical implications of these results 
for the scale dependence of the non-singlet quark distributions are 
illustrated in section 5. Except for very small values of $x$, the 
perturbation series appears to be well-behaved even down to sub-GeV 
scales where the initial distributions have been studied using 
non-perturbative methods for example in Refs.\ 
\cite{Weigel:1997kw,Diakonov:1996sr,Diakonov:1997vc,Pobylitsa:1998tk,%
Schroeder:1999fr,Weigel:1999pc}.
Finally we briefly summarize our findings in section~6.
%
%
\section{Notations and method} 
\label{sec:method}
%
%
We start by setting up our notations for the non-singlet combinations 
of parton distributions and the splitting functions governing their
evolution. The number distributions of quarks and antiquarks in a 
hadron are denoted by $q_i(x,\mu_f^{\,2})$ and $\bar{q}_i(x,
\mu_f^{\,2})$, respectively, where $x$ represents the fraction of the 
hadron momentum carried by the parton and $\mu_f$ stand for the 
factorization scale. There is no need to introduce a renormalization 
scale $\mu_r$ different from $\mu_f$ at this point. The subscript $i$ 
indicates the flavour of the (anti-)quark, with $i = 1,\ldots ,\nf\,$ 
for $\nf$ flavours of light quarks.

The general structure of the (anti-)quark (anti-)quark splitting 
functions, constrained by charge conjugation invariance and flavour 
symmetry, is given by
\bea
  P_{{\rm q}_{i}{\rm q}_{k}} \: = \: 
  P_{\bar{{\rm q}}_{i}\bar{{\rm q}}_{k}} 
  &\! =\! & \delta_{ik} P_{{\rm q}{\rm q}}^{\,\rm v} 
        + P_{{\rm q}{\rm q}}^{\,\rm s} \nonumber \\
  P_{{\rm q}_{i}\bar{{\rm q}}_{k}} \: = \: 
  P_{\bar{{\rm q}}_{i}{\rm q}_{k}} 
  &\! =\! & \delta_{ik} P_{{\rm q}\bar{{\rm q}}}^{\,\rm v} 
        + P_{{\rm q}\bar{{\rm q}}}^{\,\rm s} 
  \:\: .
\eea
In the  expansion in powers of $\as$ the flavour-diagonal (`valence') 
quantity $P_{\rm qq}^{\,\rm v}$ starts at first order, while 
$P_{{\rm q}\bar{{\rm q}}}^{\,\rm v}$ and the flavour-independent 
(`sea') contributions $P_{{\rm qq}}^{\,\rm s}$ and 
$P_{{\rm q}\bar{{\rm q}}}^{\,\rm s}$ are of order $\alpha_{\rm s}^2$. 
A non-vanishing difference $P_{{\rm qq}}^{\,\rm s} - 
P_{{\rm q}\bar{{\rm q}}}^{\,\rm s}$ occurs for the first time at the
third order. 

This general structure leads to three independently evolving types of 
non-singlet distributions: The evolution of the flavour asymmetries
\beq
\label{eq:qpm} 
  q_{{\rm ns},ik}^{\,\pm} \: = \: q_i^{} \pm 
  \bar{q}_i^{} - (q_k^{} \pm \bar{q}_k^{})
\eeq
and of linear combinations thereof, hereafter generically denoted by 
$q_{\rm ns}^{\pm}$, is governed by 
\beq
\label{eq:ppm} 
  P_{\rm ns}^{\,\pm} \: = \: P_{{\rm q}{\rm q}}^{\,\rm v}
  \pm P_{{\rm q}\bar{{\rm q}}}^{\,\rm v} \:\: .
\eeq
The sum of the valence distributions of all flavours, 
\beq
  q_{\rm ns}^{\rm v} \: = \: \sum_{r=1}^{\nf} (q_r^{} 
  - \bar{q}_r^{}) \:\: ,
\eeq
evolves with
\beq
\label{eq:pval}
  P_{\rm ns}^{\,\rm v} \: = \: P_{\rm qq}^{\,\rm v} 
  - P_{{\rm q}\bar{{\rm q}}}^{\,\rm v} + \nf (P_{\rm qq}^{\,\rm s} 
  - P_{{\rm q}\bar{{\rm q}}}^{\,\rm s}) \: \equiv \: 
  P_{\rm ns}^{\, -} + P_{\rm ns}^{\,\rm s} \:\: .
\eeq
The first moments of $P_{\rm ns}^-$ and $P_{\rm ns}^{\,\rm v}$ vanish, 
since the first moments of the distributions $q_{\rm ns}^-$ and 
$q_{\rm ns}^{\rm v}$ reflect conserved additive quantum numbers.

We expand the splitting functions in powers of $a_{\rm s}\,\equiv\,
\as /(4\pi)$, i.e.  the evolution equations for $q_{\rm ns}^{\: i}
(x,\mu_f^{\,2})$, $i=\pm,{\rm v},$ are written as 
\beq
\label{eq:evol}
  \frac{d}{d \ln \mu_f^{\,2}} \: q_{\rm ns}^{\: i}(x,\mu_f^{\,2})\: =\: 
 \sum_{n=0}\: \left( \frac{\as (\mu_f^{\,2})}{4\pi} \right)^{\! n+1}
  P_{\rm ns}^{\,(n)i} (x) \,\otimes\, q_{\rm ns}^{\, i}(x,\mu_f^{\,2}) 
\eeq
where $\otimes$ represents the standard Mellin convolution.

Our calculation is preformed in Mellin-$N$ space, i.e., we compute the
non-singlet anomalous dimensions $\gamma_{\,\rm ns}^{\,(n)i}(N)$ which 
are related by the Mellin transformation
\beq
\label{eq:Pdef}
  \gamma_{\,\rm ns}^{\:(n)i}(N) \: = \: 
  - \int_0^1 \!dx\:\, x^{\,N-1}\, P_{\,\rm ns}^{\,(n)i}(x) 
\eeq
to the splitting functions discussed above. The relative sign is the standard 
convention. Note that in the older literature an additional factor of two is 
often included in Eq.~(\ref{eq:Pdef}).

The calculation follows the methods of 
Refs.~\cite{Larin:1994vu,Larin:1997wd,Retey:2000nq,Kazakov:1988jk,Moch:1999eb}.
We employ the optical theorem and the operator product expansion to calculate 
Mellin moments of the deep-inelastic structure functions. Since we treat the 
Mellin moment $N$ as an analytical parameter, we cannot apply the techniques of 
Refs.~\cite{Larin:1994vu,Larin:1997wd,Retey:2000nq}, where the {\sc Mincer} 
program \cite{Gorishnii:1989gt,Larin:1991fz} was used as the tool to solve the 
integrals. Instead, the introduction of new techniques was necessary, and 
various aspects of those have already been discussed in 
Refs.~\cite{Moch:1999eb,Vermaseren:2000we,Moch:2002sn,Vermaseren:2002rn}.
Here we briefly summarize our approach, focussing on some parts which have not 
been presented yet.
It should be emphasized that we have at our disposal a very powerful check on 
all our derivations and calculations by letting, at any point, $N$ be some 
positive integer value. Then we can resort to the approach of 
Refs.~\cite{Larin:1994vu,Larin:1997wd,Retey:2000nq} and, with the help of the 
{\sc Mincer} program, the checking of all programs greatly simplifies.

We start by constructing the diagrams for the forward Compton reactions 
\beq
\label{eq:qaqa}
 \mbox{quark}\,(P)+\mbox{vector}\,(Q) \:\longrightarrow\: 
 \mbox{quark}\,(P)+\mbox{vector}\,(Q) \:\: ,
\end{equation}
which contribute to the non-singlet structure functions $F_2^{}$, $F_L^{}$ and 
$F_3^{}$ of deep-inelastic scattering. The $N$-th Mellin moment is given by the 
$N$-th derivative with respect to the quark momentum $P$ at \mbox{$P=0$}. 
The diagrams are generated automatically with the diagram 
generator {\sc Qgraf}~\cite{Nogueira:1991ex} and for all symbolic manipulations 
we use the latest version of 
{\sc Form}~\cite{Vermaseren:2000nd,Vermaseren:2002rp}.
The calculation is performed in dimensional regularization~\cite
{'tHooft:1972fi,Bollini:1972ui,Ashmore:1972uj,Cicuta:1972jf} with 
$D=4-2\epsilon$. The unrenormalized results in Mellin space are formulae in 
terms of the invariants determined by the colour group 
\cite{vanRitbergen:1998pn}, harmonic sums \cite{Gonzalez-Arroyo:1979df,%
Gonzalez-Arroyo:1980he,Vermaseren:1998uu,Blumlein:1998if,Moch:2001zr}
and the values $\zeta_3$, $\zeta_4$, $\zeta_5$ of the Riemann 
\mbox{$\zeta$-function}.
In physics results the terms with $\zeta_4$ cancel in $N$-space. With the help 
of an inverse Mellin transformation the results can be transformed to harmonic 
polylogarithms \cite{Goncharov,Borwein,Remiddi:1999ew} in Bjorken-$x$ space. 
Details have been discussed in Refs.~\cite{Moch:1999eb,Moch:2000zw}.
The renormalization is carried out in the \MSb-scheme 
\cite{'tHooft:1973mm,Bardeen:1978yd} as described in 
Ref.~\cite{Larin:1994vu,Larin:1997wd,Retey:2000nq}.

The complete non-singlet contributions to the structure functions can be 
obtained from three Lorentz projections of the amplitude for the process
(\ref{eq:qaqa}), that is with $g^{\mu\nu}$, $P^\mu P^\nu$ and with 
$\epsilon^{PQ\mu\nu} \equiv \epsilon^{\alpha\beta\mu\nu}P_\alpha Q_\beta$.
For the projection with $g^{\mu\nu}$ and $P^\mu P^\nu$ one has two vector-like 
couplings, whereas for the projection with $\epsilon^{PQ\mu\nu}$ one has the 
product of a vector and an axial-vector coupling. The axial nature leads to the 
need for additional renormalizations with $Z_A$, the axial renormalization, and 
with $Z_5$, the finite renormalization due to the treatment of the $\gamma_5$. 
This is all described in the literature~\cite{Larin:1991tj}.
For the anomalous dimensions we need only the divergent parts of the 
$g^{\mu\nu}$ and $\epsilon^{PQ\mu\nu}$ projections, but just as for the fixed 
moments we can also obtain the finite pieces which lead to the coefficient 
functions in N$^3$LO. The determination of the latter for $F_2^{}$ and $F_L^{}$
requires also the computation of the $P^\mu P^\nu$ projection which is still in 
progress. The results for the three-loop coefficient functions will thus be
presented in a future publication~\cite{MVV5}.

To solve the integrals we apply the following strategy
\cite{Moch:1999eb,Moch:2002sn}.
We set up a hierarchy of classes among all diagrams depending on the topology, 
for instance ladder, Benz or non-planar. Within a certain topology, we define a 
sub-hierarchy depending on the number of $P$-dependent propagators. We define 
basic building blocks (BBB's) as diagrams of a given topology in which the 
quark momentum $P$ flows only through a single line in the diagram, while 
composite building blocks (CBB's) denote all diagrams with more than one 
$P$-dependent propagator. We determine reduction schemes that map the CBB's of 
a given topology class to the BBB's of the same topology class or to simpler 
CBB topologies. Subsequently, we use reduction identities that express the 
BBB's of a given topology class in terms of BBB's of simpler topologies.

This procedure has been discussed to some extent in 
Refs.~\cite{Moch:1999eb,Moch:2002sn}. 
It exploits various categories of relations between the integrals which can be 
derived as follows. For a generic loop integral depending on external momenta 
$P$ and $Q$, the first category are integration-by-parts identities
\cite{'tHooft:1972fi,Chetyrkin:1981qh}, 
\begin{eqnarray}
\int\, \prod_n\, d^Dp_n\, {\partial \over \partial p_i^{\,\mu}}\, p_j^{\,\mu}
 \:\times\: (\dots) & = & 0\, .
  \label{eq:ibp}
\end{eqnarray}
These give a number of nontrivial relations by making various choices for the 
$p_i$ and $p_{\!j}$ from the loop momenta. Additionally $p_{\!j}$ can be equal 
to $P$ or $Q$.
The second category is based on scaling arguments~\cite{Moch:1999eb} in Mellin 
space.  They involve applying one of the operators
\beq
Q^{\,\mu} {\partial \over \partial Q^{\,\mu}}\:\: , \quad\quad\quad
P^{\,\mu} {\partial \over \partial Q^{\,\mu}}\:\: , \quad\quad\quad
P^{\,\mu} {\partial \over \partial P^{\,\mu}}
  \label{eq:scale}
\eeq
both inside the integral and to the integrated result. The scaling in Mellin 
space tells us the effect of these operators on the integrated result, while 
inside the integral we just work out the derivative. These relations naturally 
involve polynomials linear in $N$. The fourth operator of this kind,
\begin{eqnarray}
Q^\mu {\partial \over \partial P^\mu}\:\: , 
  \label{eq:scale2}
\end{eqnarray}
cannot be used naively in this context, because it does not commute with the 
limit $P \!\cdot\! P \to 0$.
More care is needed in this case and we will come back to this shortly.

A third category of relations is obtained along the lines of the 
Passarino--Veltman decomposition into form factors~\cite{Passarino:1979jh}. 
In Mellin space we write 
\begin{eqnarray}
   \int\, \prod_n\, d^Dp_n\, p_i^{\,\mu}\, \times\, (\dots) & = &
   Q^{\,\mu}\, I_Q + P^{\,\mu}\, I_P\, \:\: ,
\label{eq:formfac}
\end{eqnarray}
where $I_Q$ and $I_P$ are the two form factors. By contracting 
Eq.~(\ref{eq:formfac}) either with $Q_\mu$ or $P_\mu$, the $I_Q$ and $I_P$ are 
determined in terms of a number of integrals.  Next, by taking the derivative 
with respect to $Q_\mu$, the relevant identities can be obtained.
Because the momentum $p_i$ can be any of the loop momenta, 
Eq.~(\ref{eq:formfac}) gives us as many relations as there are loops. Again, in 
Mellin space, these relations contain polynomials linear in $N$.

The fourth and the fifth category of relations are new. Together with the form 
factor relations from Eq.~(\ref{eq:formfac}) they were crucial in setting up 
the reduction scheme for the three-loop topologies. They are based on operators 
that do not commute with the limit $P \!\cdot\! P \to 0$. 
In the fourth category, one considers the dimensionless operators
\begin{eqnarray}
  O_1  & = & {P \cdot Q \over Q \cdot Q}\: 
  Q^\mu {\partial \over \partial P^\mu}\, , 
\label{eq:Veqs1}
\\
  O_2  & = & {P \cdot Q} \:\: {\partial \over \partial P^\mu}\, 
  {\partial \over \partial Q^\mu}\, ,   
\label{eq:Veqs2}
\\
  O_3  & = & {(P \cdot Q)^2 \over Q \cdot Q}\: 
  {\partial \over \partial P^\mu}\, {\partial \over \partial P^\mu} \:\: .
\label{eq:Veqs3}
\end{eqnarray}
Each individual operator $O_i$ does not commute with the limit $P\cdot P\to 0$, 
but certain linear combinations of the $O_i$ do.  However, one has to extend 
the ansatz based on scaling arguments in $N$-space. Specifically, one has for 
the $N$-th moment of an integral $I(N)$ 
\begin{eqnarray}
        I(N) & = & 
\bigg({2 P \cdot Q \over Q \cdot Q}\bigg)^N \, (Q \cdot Q)^\alpha\, C^{(0)}_N 
+
\bigg({2 P \cdot Q \over Q \cdot Q}\bigg)^{N-2}\, {P \cdot P \over Q \cdot Q}\, 
(Q \cdot Q)^\alpha\,  C^{(2)}_N \,+\, \dots \:\:  ,
\label{eq:scaleansatz}
\end{eqnarray}
where the $C^{(0)}_N$ and $C^{(2)}_N$ are dimensionless functions of $N$, and 
$\alpha$ adjusts the mass dimensions. The novel feature is here the 
term $C^{(2)}_N$ proportional to $P \cdot P$, which one may call higher twist.
In contrast, for the relations based on Eq.~(\ref{eq:scale}) it was sufficient 
to restrict the ansatz to $C^{(0)}_N$.

Applying the differential operators $O_i$ in Eqs.~(\ref{eq:Veqs1}) -- 
(\ref{eq:Veqs3}) to the ansatz (\ref{eq:scaleansatz}), one finds that the 
combinations 
\begin{eqnarray}
         2 (\alpha +1-N) O_1 - O_2\:\: , \quad\quad\quad 
         (2N-4+D) O_1 - O_3 
\label{eq:Vops}
\end{eqnarray}
do commute with the limit $P\!\cdot\! P\to 0$. That is to say, any dependence 
on the higher twist term $C^{(2)}_N$ vanishes in this limit and one is left 
with only contributions from $C^{(0)}_N$. Eq.~(\ref{eq:Vops}) adds two more 
relations, which in Mellin space contain quadratic polynomials in $N$ due to 
the differential operators of second order.
We have checked that differential operators of yet a higher order in $P$ and 
$Q$ do not add any new information.

Finally, the fifth category of relations again uses the form factor approach of 
Eq.~(\ref{eq:formfac}). However, now we do not take the derivative with respect 
to $Q_\mu$ but with respect to $P_\mu$. Some extra book-keeping is needed here, 
since one has to take along terms proportional to $P \!\cdot\! P$.
Let us write Eq.~(\ref{eq:formfac}) as 
\bea
  p_i^{\,\mu}\, I & = & Q^{\,\mu}\, I_Q + P^{\,\mu}\, I_P \:\: .
\label{eq:Pformfac}
\eea
Taking the derivative of Eq.~(\ref{eq:Pformfac}) with respect to $P_\mu$ in 
$N$-space one finds 
\begin{eqnarray}
   {\partial \over \partial P^\mu}\, p_i^{\,\mu}\, I & = &
   Q^{\,\mu}\, {\partial \over \partial P^{\,\mu}}\, I_Q
   + (D+N-1)\, I_P \:\: .
\label{eq:partialPformfac}
\end{eqnarray}
Solving Eq.~(\ref{eq:Pformfac}) for $I_Q$ and $I_P$ as above, however keeping 
all terms $P \!\cdot\! P$, substituting into Eq.~(\ref{eq:partialPformfac}) 
and finally taking the limit $P \!\cdot\! P \to 0$, we find
\begin{eqnarray}
  {\partial \over \partial P^\mu}\, p_i^{\,\mu}\, I & = & 
    {P \cdot p_i \over P \cdot Q}\: Q^{\,\mu}\, 
    {\partial \over \partial P^{\,\mu}}\, I
  + (D+N-2)\, {Q\cdot p_i\,  P \cdot Q 
  - Q \cdot Q\,  P \cdot p_i \over (P \cdot Q)^2 }\, I \:\: .
\label{eq:finPformfac}
\end{eqnarray}
Again, as the momentum $p_i$ can be any of the loop momenta, 
Eq.~(\ref{eq:finPformfac}) gives us as many relations with polynomials linear 
in $N$ as there are loops.

Taken together, the reductions of category one to five suffice to obtain a 
complete reduction scheme. In particular, the reduction equations of category 
two to five involve explicitly the parameter $N$ of the Mellin moment. They 
give rise to difference equations in $N$ for an integral $I(N)$,
\beq
\label{eq:deq}
  a_0(N)\, I(N) + a_1(N)\,  I(N-1) + \ldots + a_m(N)\,  I(N-m) \: = \: G(N)
  \:\: ,
\eeq
in which the function $G$ refers to a combination of integrals of simpler 
topologies. Zeroth order equations are of course trivial, although sometimes
the function $G$ can contain thousands of terms. First order difference 
equations can be solved analytically in a closed form, introducing one sum.
Higher order difference equations on the other hand can be solved 
constructively, sometimes with considerable effort, by making an ansatz for the 
solution in terms of harmonic sums. 
For the present calculation we had to go up to fourth order for certain types 
of integrals. 

Due to the difference equations, which have to be solved in a successive way, 
a strict hierarchy for topology classes is introduced in the reduction scheme. 
For a given integral $I$, a difference equation as in Eq.~(\ref{eq:deq}),  
with some (often lenghty) function $G$ expressed in terms of harmonic sums, 
can be solved in terms of harmonic sums again. 
Subsequently, the result for $I$ can be part of the inhomogenous term in a 
difference equation for another, more complicated integral. This requires the 
tabulation of a large number CBB and BBB integrals, because each integral is 
typically used many times, thus it saves computer time and disk space. 
Only this tabulation, which required the addition of features to {\sc Form} 
\cite{Vermaseren:2002rp}, renders the calculation feasible with current
computing resources. For the complete project, including Refs.~\cite{MVV4,MVV5},
we have collected tablebases with more than $100.000$ integrals and a total 
size of tables of more than 3 GByte.
%
%
\setcounter{equation}{0}
\section{Results in Mellin space}
\label{sec:results}
%
%
Here we present the anomalous dimensions $\gamma^{\,\pm,{\rm s}}
_{\,\rm ns}(N)$ in the \MSb-scheme up to the third order in the 
running coupling constant $\as$, expanded in powers of $\as /(4\pi)$. 
These quantities can be expressed in terms of harmonic 
sums~\cite{Gonzalez-Arroyo:1979df,Gonzalez-Arroyo:1980he,%
Vermaseren:1998uu,Blumlein:1998if}. Following the notation of 
Ref.~\cite{Vermaseren:1998uu}, these sums are recursively defined by
\beq
\label{eq:Hsum1}
  S_{\pm m}(M) \: = \: \sum_{i=1}^{M}\: \frac{(\pm 1)^m}{i^{\, m}}
\eeq
and 
\beq
\label{eq:Hsum2}
  S_{\pm m_1,m_2,\ldots,m_k}(M) \: = \: \sum_{i=1}^{M}\: 
  \frac{(\pm 1)^{m_1}}{i^{\, m_1}}\: S_{m_2,\ldots,m_k}(i) \:\: .
\eeq
The sum of the absolute values of the indices $m_k$ defines the weight
of the harmonic sum. In the $n$-loop anomalous dimensions written down
below one encounters sums up to weight $2n-1$.
 
In order to arrive at a reasonably compact representation of our 
results, we employ the abbreviation $S_{\vec{m}}\,\equiv\, 
S_{\vec{m}}(N)$ in what follows, together with the notation
\beq
\label{eq:shiftN}
  \Npm \, S_{\vec{m}} \: = \: S_{\vec{m}}(N \pm 1) \:\: , \quad\quad
  \Npmi\, S_{\vec{m}} \: = \: S_{\vec{m}}(N \pm i) 
\eeq
for arguments shifted by $\pm 1$ or a larger integer $i$. In this
notation the well-known one-loop (LO) anomalous dimension 
\cite{Gross:1973rr,Georgi:1974sr} reads
\bea
  \gamma^{\,(0)}_{\,\rm ns}(N) & \! = \! &
         \colour4colour{\cf} \* \big(
            2 \* (\Nminus+\Nplus) \* \S(1)
          - 3
          \big)
\:\ ,\label{eq:gqq0}
\eea
and the corresponding two second-order (NLO) non-singlet quantities 
\cite{Floratos:1977au,Gonzalez-Arroyo:1979df} are given by
\bea
  &&\gamma^{\,(1)+}_{\,\rm ns}(N) \:\: = \:
  4\, \* \colour4colour{\ca \* \cf} \* \bigg(
            2\, \* \Nplus \* \S(3)
          - {17 \over 24}
          - 2 \* \S(-3)
          - {28 \over 3} \* \S(1)
          + (\Nminus+\Nplus) \* \bigg[
            {151 \over 18} \* \S(1)
          + 2 \* \Ss(1,-2)
          - {11 \over 6} \* \S(2)
          \bigg]
          \bigg)
  \nonumber\\&& \mbox{} \quad
+ 4\, \* \colour4colour{\cf \* \nf} \* \bigg(
            {1 \over 12}
          + {4 \over 3} \* \S(1)
          - (\Nminus+\Nplus) \* \bigg[
            {11 \over 9} \* \S(1)
          - {1 \over 3} \* \S(2)
          \bigg]
          \bigg)
+ 4\, \* \colour4colour{\cf^{2}} \* \bigg(
            4 \* \S(-3)
          + 2 \* \S(1)
          + 2 \* \S(2)
          - {3 \over 8}
  \nonumber\\&& \mbox{} \quad
          + \Nminus \* \bigg[
            \S(2)
          + 2 \* \S(3)
          \bigg]
          - (\Nminus+\Nplus) \* \bigg[
            \S(1)
          + 4 \* \Ss(1,-2)
          + 2 \* \Ss(1,2)
          + 2 \* \Ss(2,1)
          + \S(3)
          \bigg]
          \bigg)
 \:\: ,\label{eq:gqq1p}
 \\[2mm] 
  &&\gamma^{\,(1)-}_{\,\rm ns}(N) \:\: = \:\:
     \gamma^{\,(1)+}_{\,\rm ns}(N)
       + 16\, \*  \colour4colour{\cf \* \bigg(\cf - {\ca \over 2} \bigg)}  \*  \bigg(
            (\Nminus-\Nplus) \* \bigg[
            \S(2)
          - \S(3)
          \bigg]
          - 2 \* (\Nminus+\Nplus-2) \* \S(1)
          \bigg)
 \:\: . \quad \nonumber\\&& \mbox{}
 \label{eq:gqq1m}
\eea

The three-loop (NNLO, N$^{\,2}$LO) contribution to the anomalous 
dimension $\gamma^{\,+}_{\,\rm ns}(N)$ corresponding to the upper sign 
in Eq.~(\ref{eq:ppm}) reads
\bea
  &&\gamma^{\,(2)+}_{\,\rm ns}(N) \:\: = \:\: 
 16\, \* \colour4colour{\ca \* \cf \* \nf} \* \bigg( 
            {3 \over 2} \* \z3
          - {5 \over 4}
          + {10 \over 9} \* \S(-3)
          - {10 \over 9} \* \S(3)
          + {4 \over 3} \* \Ss(1,-2)
          - {2 \over 3} \* \S(-4)
          + 2 \* \Ss(1,1)
          - {25 \over 9} \* \S(2)
  \nonumber\\&& \mbox{}
          + {257 \over 27} \* \S(1)
          - {2 \over 3} \* \Ss(-3,1)
       - \Nplus \*  \bigg[
            \Ss(2,1)
          - {2 \over 3} \* \Ss(3,1)
          - {2 \over 3} \* \S(4)
          \bigg]
          - (\Nplus-1)  \*  \bigg[
            {23 \over 18} \* \S(3)
          - \S(2)
         \bigg]
       - (\Nminus+\Nplus)  \*   \bigg[
            \Ss(1,1)
  \nonumber\\&& \mbox{}
          + {1237 \over 216} \* \S(1)
          + {11 \over 18} \* \S(3)
          - {317 \over 108} \* \S(2)
          + {16 \over 9} \* \Ss(1,-2)
          - {2 \over 3} \* \Sss(1,-2,1)
          - {1 \over 3} \* \Ss(1,-3)
          - {1 \over 2} \* \Ss(1,3)
          - {1 \over 2} \* \Ss(2,1)
          - {1 \over 3} \* \Ss(2,-2)
          + \S(1) \* \z3
  \nonumber\\&& \mbox{}
          + {1 \over 2} \* \Ss(3,1)
           \bigg]
          \bigg)
       + 16\, \* \colour4colour{\cf \* \ca^2}  \*  \bigg(
            {1657 \over 576}
          - {15 \over 4} \* \z3
          + 2 \* \S(-5)
          + {31 \over 6} \* \S(-4)
          - 4 \* \Ss(-4,1)
          - {67 \over 9} \* \S(-3)
          + 2 \* \Ss(-3,-2)
  \nonumber\\&& \mbox{}
          + {11 \over 3} \* \Ss(-3,1)
          + {3 \over 2} \* \S(-2)
          - 6 \* \S(-2) \* \z3
          - 2 \* \Ss(-2,-3)
          + 3 \* \Ss(-2,-2)
          - 4 \* \Sss(-2,-2,1)
          + 8 \* \Sss(-2,1,-2)
          - {1883 \over 54} \* \S(1)
  \nonumber\\&& \mbox{}
          - 10 \* \Ss(1,-3)
          - {16 \over 3} \* \Ss(1,-2)
          + 12 \* \Sss(1,-2,1)
          + 4 \* \Ss(1,3)
          - 4 \* \Ss(2,-2)
          - {5 \over 2} \* \S(4)
          + {1 \over 2} \* \S(5)
          + {176 \over 9} \* \S(2)
          + {13 \over 3} \* \S(3)
  \nonumber\\&& \mbox{}
          + (\Nminus+\Nplus-2) \* \bigg[ 
 	    3 \* \S(1) \* \z3
          + 11 \* \Ss(1,1)
          - 4 \* \Sss(1,1,-2)
	\bigg]
          + (\Nminus+\Nplus) \* \bigg[
            {9737 \over 432} \* \S(1)
          - 3 \* \Ss(1,-4) 
          + {19 \over 6} \* \Ss(1,-3)
  \nonumber\\&& \mbox{}
          + 8 \* \Sss(1,-3,1)
          + {91 \over 9} \* \Ss(1,-2)
          - 6 \* \Ss(1,-2,-2)
          - {29 \over 3} \* \Sss(1,-2,1)
          + 8 \* \Sss(1,1,-3)
          - 16 \* \Ssss(1,1,-2,1)
          - 4 \* \Sss(1,1,3)
          - {19 \over 4} \* \Ss(1,3)
  \nonumber\\&& \mbox{}
          + 4 \* \Sss(1,3,1)
          + 3 \* \Ss(1,4)
          + 8 \* \Sss(2,-2,1)
          + 2 \* \Ss(2,3)
          - \Ss(3,-2)
          + {11 \over 12} \*\Ss(3,1)
          - \Ss(4,1)
          - 4 \* \Ss(2,-3)
          +  {1 \over 6} \* \Ss(2,-2)
          - {1967 \over 216} \* \S(2)
  \nonumber\\&& \mbox{}
          + {121 \over 72} \* \S(3)
	\bigg]
          - (\Nminus-\Nplus) \* \bigg[ 
            3 \* \S(2) \* \z3
          + 7 \* \Ss(2,1)
          - 3 \* \Sss(2,1,-2)
          + 2 \* \Sss(2,-2,1)
          - {1 \over 4} \* \Ss(2,3)
          - {3 \over 2} \* \Ss(3,-2)
          - {29 \over 6} \*\Ss(3,1)
  \nonumber\\&& \mbox{}
          + {11 \over 4} \* \Ss(4,1)
          + {1 \over 2} \* \Ss(2,-3)
          - \Ss(2,-2)
	\bigg]
          + \Nplus \*  \bigg[
            {28 \over 9} \* \S(3)
          - {2376 \over 216} \* \S(2)
          - {8 \over 3} \* \S(4)
          - {5 \over 2} \* \S(5)
          \bigg]
          \bigg)
+ 16\, \* \colour4colour{\cf \* \n2f} \* \bigg( 
            {17 \over 144}
  \nonumber\\&& \mbox{}
          - {13 \over 27} \* \S(1)
          + {2 \over 9} \* \S(2)
              + (\Nminus+\Nplus) \*  \bigg[
            {2 \over 9} \* \S(1)
          - {11 \over 54} \* \S(2)
          + {1 \over 18} \* \S(3)
          \bigg]
          \bigg)
       + 16\, \* \colour4colour{\cf^2 \* \ca}  \*  \bigg(
            {45 \over 4} \* \z3
          - {151 \over 64}
          - 10 \* \S(-5)
  \nonumber\\&& \mbox{}
          - {89 \over 6} \* \S(-4)
          + 20 \* \Ss(-4,1)
          + {134 \over 9} \* \S(-3)
          - 2 \* \Ss(-3,-2)
          - {31 \over 3} \* \Ss(-3,1)
          + 2 \* \Ss(-3,2)
          - {9 \over 2} \* \S(-2)
          + 18 \* \S(-2) \* \z3
          + 10 \* \Ss(-2,-3)
  \nonumber\\&& \mbox{}
          - 6 \* \Ss(-2,-2)
          + 8 \* \Sss(-2,-2,1)
          - 28 \* \Sss(-2,1,-2)
          + 46 \* \Ss(1,-3)
          + {26 \over 3} \* \Ss(1,-2)
          - 48 \* \Sss(1,-2,1)
          + {28 \over 3} \* \Ss(1,2)
          - {185 \over 6} \* \S(3)
  \nonumber\\&& \mbox{}
          - 8 \* \Ss(1,3)
          + 2 \* \Ss(3,-2)
          - 4 \* \S(5)
          - (\Nminus+\Nplus-2) \* \bigg[
            9 \* \S(1) \* \z3
          - {133 \over 36} \* \S(1)
          + {209 \over 6} \* \Ss(1,1)
          - 14 \* \Sss(1,1,-2)
          - {242 \over 18} \* \S(2)
  \nonumber\\&& \mbox{}
          + 9 \* \Ss(2,-2)
          + {33 \over 4} \* \S(4)
          - 3 \* \Ss(3,1)
          + {14 \over 3} \* \Ss(2,1)
	\bigg]
          + (\Nminus+\Nplus) \* \bigg[
            17 \* \Ss(1,-4)
          - {107 \over 6} \* \Ss(1,-3)
          - 32 \* \Sss(1,-3,1)
  \nonumber\\&& \mbox{}
          - {173 \over 9} \* \Ss(1,-2)
          + 16 \* \Sss(1,-2,-2)
          + {103 \over 3} \* \Sss(1,-2,1)
          - 2 \* \Sss(1,-2,2)
          - 36 \* \Sss(1,1,-3)
          + 56 \* \Ssss(1,1,-2,1)
          + 8 \* \Sss(1,1,3)
  \nonumber\\&& \mbox{}
          - {109 \over 9} \* \Ss(1,2)
          - 4 \* \Sss(1,2,-2)
          + {43 \over 3} \* \Ss(1,3)
          - 8 \* \Sss(1,3,1)
          - 11 \* \Ss(1,4)
          + {11 \over 3} \* \Ss(2,2)
          + 21 \* \Ss(2,-3)
          - 30 \* \Sss(2,-2,1)
          - 4 \* \Sss(2,1,-2)
  \nonumber\\&& \mbox{}
          - 5 \* \Ss(2,3)
          - \Ss(4,1)
          + {31 \over 6} \* \Ss(2,-2)
          - {67 \over 9} \* \Ss(2,1)
	\bigg]
          + (\Nminus-\Nplus) \* \bigg[
            9 \* \S(2) \* \z3
          + 2 \* \Ss(2,-3)
          + 4 \* \Sss(2,-2,1)
          - 12 \* \Sss(2,1,-2)
  \nonumber\\&& \mbox{}
          - 2 \* \Ss(2,3)
          + 13 \* \Ss(4,1)
          + {1 \over 2} \* \Ss(2,-2)
          + {11 \over 2} \* \S(4)
          - {33 \over 2} \* \Ss(3,1)
          + {59 \over 9} \* \S(3)
          + {127 \over 6} \* \Ss(2,1)
          - {1153 \over 72} \* \S(2)
	\bigg]
          + \Nplus \* \bigg[
            8 \* \Ss(3,-2)
  \nonumber\\&& \mbox{}
          + {4 \over 3} \* \Ss(3,1)
          - 2 \* \Ss(3,2)
          + 14 \* \S(5)
          + {23 \over 6} \* \S(4)
          + {73 \over 3} \* \S(3)
          + {151 \over 24} \* \S(2)
	\bigg]
          \bigg)
+  16\, \* \colour4colour{\cf^{2} \* \nf} \* \bigg( 
            {23 \over 16}
          - {3 \over 2} \* \z3
          + {4 \over 3} \* \Ss(-3,1)
          - {59 \over 36} \* \S(2)
  \nonumber\\&& \mbox{}
          + {4 \over 3} \* \S(-4)
          - {20 \over 9} \* \S(-3)
          + {20 \over 9} \* \S(1)
          - {8 \over 3} \* \Ss(1,-2)
          - {8 \over 3} \* \Ss(1,1)
          - {4 \over 3} \* \Ss(1,2)
       + \Nplus \*  \bigg[
            {25 \over 9} \* \S(3)
          - {4 \over 3} \* \Ss(3,1)
          - {1 \over 3} \* \S(4)
          \bigg]
  \nonumber\\&& \mbox{}
               - (\Nplus-1)  \*  \bigg[
            {67 \over 36} \* \S(2)
          - {4 \over 3} \* \Ss(2,1)
          + {4 \over 3} \* \S(3)
          \bigg]
       + (\Nminus+\Nplus)  \*  \bigg[
            \S(1) \* \z3
          - {325 \over 144} \* \S(1)
          - {2 \over 3} \* \Ss(1,-3)
          + {32 \over 9} \* \Ss(1,-2)
  \nonumber\\&& \mbox{}
          - {4 \over 3} \* \Sss(1,-2,1)
          + {4 \over 3} \* \Ss(1,1)
          + {16 \over 9} \* \Ss(1,2)
          - {4 \over 3} \* \Ss(1,3)
          + {11 \over 18} \* \S(2)
          - {2 \over 3} \* \Ss(2,-2)
          + {10 \over 9} \* \Ss(2,1)
          + {1 \over 2} \* \S(4)
          - {2 \over 3} \* \Ss(2,2)
          - {8 \over 9} \* \S(3)
           \bigg]
          \bigg)
  \nonumber\\&& \mbox{}
       + 16\, \* \colour4colour{\cf^{3}}  \*  \bigg(
            12 \* \S(-5)
          - {29 \over 32}
          - {15 \over 2} \* \z3
          + 9 \* \S(-4)
          - 24 \* \Ss(-4,1)
          - 4 \* \Ss(-3,-2)
          + 6 \* \Ss(-3,1)
          - 4 \* \Ss(-3,2)
          + 3 \* \S(-2)
          + 25 \* \S(3)
  \nonumber\\&& \mbox{}
          - 12 \* \S(-2) \* \z3
          - 12 \* \Ss(-2,-3)
          + 24 \* \Sss(-2,1,-2)
          - 52 \* \Ss(1,-3)
          + 4 \* \Ss(1,-2)
          + 48 \* \Sss(1,-2,1)
          - 4 \* \Ss(3,-2)
          + {67 \over 2} \* \S(2)
          - 17 \* \S(4)
  \nonumber\\&& \mbox{}
          + (\Nminus+\Nplus-2) \*  \bigg[
            6 \* \S(1) \* \z3
          - {31 \over 8} \* \S(1)
          + 35 \* \Ss(1,1)
          - 12 \* \Sss(1,1,-2)
          + \Ss(1,2)
          + 10 \* \Ss(2,-2)
          + \Ss(2,1)
          + 2 \* \Ss(2,2)
          - 2 \* \Ss(3,1)
  \nonumber\\&& \mbox{}
          - 3 \* \S(5)
          \bigg]
          + (\Nminus+\Nplus) \*  \bigg[
            23 \* \Ss(1,-3)
          - 22 \* \Ss(1,-4)
          + 32 \* \Sss(1,-3,1)
          - 2 \* \Ss(1,-2)
          - 8 \* \Sss(1,-2,-2)
          - 30 \* \Sss(1,-2,1)
          - 6 \* \Ss(1,3)
  \nonumber\\&& \mbox{}
          + 4 \* \Sss(1,-2,2)
          + 40 \* \Sss(1,1,-3)
          - 48 \* \Ssss(1,1,-2,1)
          + 8 \* \Sss(1,2,-2)
          + 4 \* \Sss(1,2,2)
          + 8 \* \Sss(1,3,1)
          + 4 \* \Ss(1,4)
          + 28 \* \Sss(2,-2,1)
          + 4 \* \Ss(2,1,2)
  \nonumber\\&& \mbox{}
          + 4 \* \Sss(2,2,1)
          + 4 \* \Sss(3,1,1)
          - 4 \* \Ss(3,2)
          + 8 \* \Sss(2,1,-2)
          - 26 \* \Ss(2,-3)
          - 2 \* \Ss(2,3)
          - 4 \* \Ss(3,-2)
          - 3 \* \Ss(2,-2)
          - 3\* \Ss(2,2)
          + {3 \over 2} \* \S(4)
          \bigg]
  \nonumber\\&& \mbox{}
          + (\Nminus-\Nplus) \*  \bigg[
            12 \* \Sss(2,1,-2)
          - 6 \* \S(2) \* \z3
          - 2 \* \Ss(2,-3)
          + 3 \* \Ss(2,3)
          + 2 \* \Ss(3,-2)
          - {81 \over 4} \* \Ss(2,1)
          + 14 \* \Ss(3,1)
          - 5 \* \Ss(2,-2)
  \nonumber\\&& \mbox{}
          - {1 \over 2} \* \Ss(2,2)
          + {15 \over 8} \* \S(2)
          + {1 \over 2} \* \S(3)
          - 13 \* \Ss(4,1)
          + 4 \* \S(5)
          \bigg]
          + \Nplus \*  \bigg[
            14 \* \S(4)
          - {265 \over 8} \* \S(2)
          - {87 \over 4} \* \S(3)
          - 4 \* \Ss(4,1)
          - 4 \* \S(5)
          \bigg]
          \bigg)
\: \: .\label{eq:gqq2p}
\eea
The third-order result for the anomalous dimension $\gamma^{\,-}_
{\,\rm ns}(N)$ corresponding to the lower sign in Eq.~(\ref{eq:ppm}) 
is given by
\bea
  &&\gamma^{\,(2)-}_{\,\rm ns}(N) \:\: = \:\:  
	\gamma^{\,(2)+}_{\,\rm ns}(N)
 + 16\, \* \colour4colour{\ca \* \cf \* \bigg(\cf - {\ca \over 2} \bigg)} \* \bigg( 
          (\Nminus+\Nplus-2) \*  \bigg[
            {367 \over 18}   \* \S(1) 
          + 12 \* \S(1) \* \z3
          + 2 \* \Ss(1,-2) 
  \nonumber\\&& \mbox{}
          + 4 \* \Ss(1,-3)
          + 8 \* \Sss(1,-2,1) 
          + {140 \over 3} \* \Ss(1,1)
          - 16 \* \Sss(1,1,-2) 
          - \S(5)
          - 8 \* \Ss(3,1)
          - \S(4)
          \bigg]
          + (\Nminus-\Nplus) \*  \bigg[
            4 \* \S(5) 
          - 12 \* \S(2) \* \z3
  \nonumber\\&& \mbox{}
          - 4 \* \Ss(2,-3)
          - 8 \* \Sss(2,-2,1)
          - {70 \over 3} \* \Ss(2,1) 
          + 16 \* \Sss(2,1,-2) 
          + 4 \* \Ss(3,-2)
          - 8 \* \Ss(4,1)
          + {70 \over 3} \* \Ss(3,1) 
          + {13 \over 3} \* \S(4) 
          - {41 \over 18} \* \S(2)
  \nonumber\\&& \mbox{}
          + 2 \* \Ss(2,-2) 
          - {152 \over 9} \* \S(3) 
          \bigg]
          + 4 \* (\Nplus-1) \* \bigg[
            4  \* \Ss(2,-2)
          - 8 \* \S(2)
          - \S(3)
          \bigg]
          \bigg)
  + 16 \* \colour4colour{\cf \* \nf \* \bigg(\cf - {\ca \over 2} \bigg)}
  \nonumber\\&& \mbox{}
          \cdot \bigg( (\Nminus+\Nplus-2) 
            \*  \bigg[ {61 \over 9} \* \S(1)
          - {8 \over 3}  \* \Ss(1,1) 
          \bigg]
          + (\Nminus-\Nplus) \*  \bigg[
            {4 \over 3} \* \Ss(2,1) 
          - {41 \over 9} \* \S(2) 
          + {38 \over 9} \* \S(3) 
          - {4 \over 3} \* \Ss(3,1) 
          - {4 \over 3} \* \S(4) 
          \bigg]
          \bigg)
  \nonumber\\&& \mbox{}
  + 16 \* \colour4colour{\cf^{2} \* \bigg(\cf - {\ca \over 2} \bigg)}
          \* \bigg( (\Nminus+\Nplus-2) \*  \bigg[
            8 \* \Ss(1,-2) 
          - 15 \* \S(1) 
          - 12 \* \S(1) \* \z3
          - 12 \* \Ss(1,-3) 
          - 60 \* \Ss(1,1) 
  \nonumber\\&& \mbox{}
          + 24 \* \Sss(1,1,-2) 
          + 8 \* \Ss(1,2)
          + 40 \* \S(2) 
          - 12 \* \Ss(2,-2)
          + 8 \* \Ss(2,1)
          + 7 \* \S(3)
          + 12 \* \Ss(3,1)
          + 6  \* \S(5)
          \bigg]
           + (\Nminus-\Nplus) \*  \bigg[
            12 \* \S(2) \* \z3
  \nonumber\\&& \mbox{}
          - 24 \* \S(2)
          + 12 \* \Ss(2,-3) 
          + 8 \* \Ss(2,-2) 
          + 30 \* \Ss(2,1) 
          - 24 \* \Sss(2,1,-2) 
          - 4 \* \Ss(2,2) 
          - 15 \* \S(3) 
          - 38 \* \Ss(3,1) 
          + 4 \* \Ss(3,2) 
          + 24 \* \Ss(4,1) 
  \nonumber\\&& \mbox{}
          - 12 \* \S(5)
           \bigg]
  	  - (\Nplus-1) \*  \bigg[
            8 \* \Ss(3,-2) 
          + 26 \* \S(4) 
          \bigg]
          \bigg)
 \:\: . \label{eq:gqq2m}
\eea
Finally the quantity $\gamma^{\:\rm s}_{\,\rm ns}(N)$ corresponding to 
the last term in Eq.~(\ref{eq:pval}) starts at three loops with
\bea
  &&\gamma^{\,(2){\rm s}}_{\,\rm ns}(N) \:\: = \:\:  
         16\, \*  \colour4colour{\nf \, \* \dabc2n}  \*  \bigg(
            (\Nminus+\Nplus) \*  \bigg[
            {25 \over 3} \* \S(1)
          + {11 \over 12} \* \Ss(1,-3)
          - {5 \over 3} \* \Sss(1,-2,1)
          - {1 \over 6} \* \Sss(1,1,-2)
          \bigg]
  \nonumber\\&& \mbox{}
          + (\Nminus+\Nplus-2) \*  \bigg[
            {13 \over 12} \* \Ss(1,-2)
          + {91 \over 24} \* \Ss(1,1)
          - {3 \over 8} \* \Ss(1,3)
          - {1 \over 4} \* \Ss(2,-2)
          - {91 \over 48} \* \S(2)
          + {3 \over 16} \* \S(3)
          + {5 \over 8} \* \Ss(3,1)
          \bigg]
  \nonumber\\&& \mbox{}
          + {2 \over 3} \* (\Nplus-\Nplustwo)\* \bigg[
	    \S(4)
          + \Ss(2,-2)
	  - \Ss(3,1)
          \bigg]
          - {2 \over 3} \* (\Nminustwo+\Nplustwo) \*  \bigg[
            \Ss(1,-3)
          - \Sss(1,-2,1)
          - \Sss(1,1,-2)
          \bigg]
  \nonumber\\&& \mbox{}
          + (\Nminus-1) \*  \bigg[
            {1 \over 4} \* \S(4)
          + {1 \over 2} \* \S(5)
          \bigg]
          + (\Nminus-\Nplus) \*  \bigg[
            {1 \over 2} \* \Ss(2,-3)
          + {1 \over 2} \* \Ss(2,-2)
          - {109 \over 48} \* \S(2)
          - {41 \over 24} \* \Ss(2,1)
          + {67 \over 48} \* \S(3)
  \nonumber\\&& \mbox{}
          - {1 \over 2} \* \Ss(3,1)
          - \Sss(2,1,-2)
          + {1 \over 4} \* \Ss(2,3)
          + {1 \over 2} \* \Ss(3,-2)
          - {3 \over 4} \* \Ss(4,1)
          \bigg]
          - {50 \over 3} \* \S(1)
          - {1 \over 2} \* \Ss(1,-3)
          + 2 \* \Sss(1,-2,1)
          - \Sss(1,1,-2)
          \bigg)
 \:\: . \quad \label{eq:gqq2s}
\eea
 
Eqs.~(\ref{eq:gqq2p}) -- (\ref{eq:gqq2s}) represent new results of 
this article, with the exception of the (identical) $\n2f$ 
parts of Eqs.~(\ref{eq:gqq2p}) and (\ref{eq:gqq2m}) which have been 
obtained by Gracey in Ref.~\cite{Gracey:1994nn} and of the 
contribution linear in $\nf$ in Eq.~(\ref{eq:gqq2p}) which we have 
published before~\cite{Moch:2002sn}.
All our results agree with the fixed moments determined before using 
the {\sc Mincer} program~\cite{Gorishnii:1989gt,Larin:1991fz}, i.e. 
Eq.~(\ref{eq:gqq2p}) reproduces the even moments $N=2,\dots,14$ 
computed in Refs.~\cite{Larin:1994vu,Larin:1997wd,Retey:2000nq}, while 
Eqs.~(\ref{eq:gqq2m}) and (\ref{eq:gqq2s}) reproduce the odd moments 
$N=1,\dots,13$ also obtained in Ref.~\cite{Retey:2000nq}.

\begin{figure}[tbh]
\label{pic:gam-plus}
\centerline{\epsfig{file=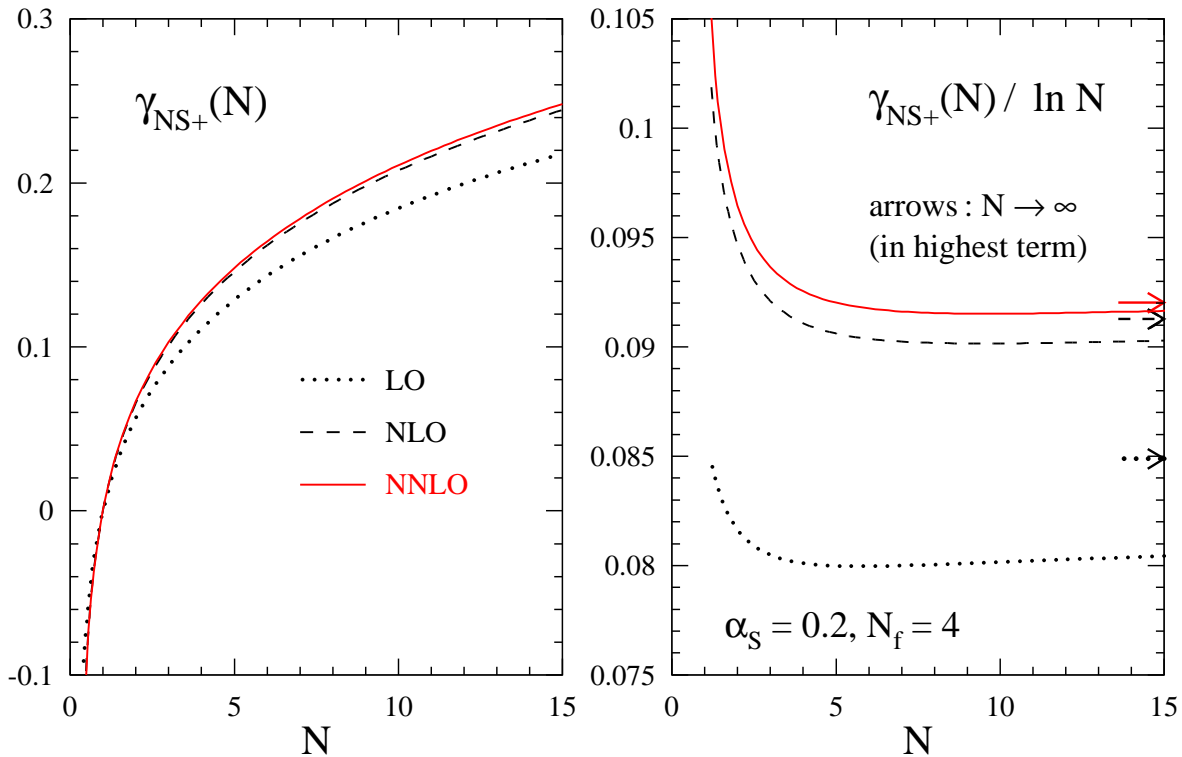,width=16cm,angle=0}}
\vspace{-1mm}
\caption{The perturbative expansion of the anomalous dimension 
 $\gamma_{\,\rm ns}^{\, +}(N)$ for four flavours at $\as = 0.2$. 
 In the right part the leading $N$-dependence for large $N$ has been
 divided out, and the corresponding asymptotic limits are indicated as 
 discussed in the text.}
\vspace*{2mm}
\end{figure}

The results (\ref{eq:gqq0}), (\ref{eq:gqq1p}) and (\ref{eq:gqq2p}) for
$\gamma_{\rm ns}^{\, +}(N)$ are assembled in Fig.~1 for four active
flavours and a typical value $\as = 0.2$ for the strong coupling
constant (recall that the terms up to order $\alpha_{\rm s}^{\, n+1}$
are included at N$^{\,\rm n}$LO). Numerically, the colour factors take 
the values $\cf=4/3, \ca=3$ and ${d^{abc}d_{abc}}/n_c=40/9$.
Note that the latter normalization is different from that employed in
Ref.~\cite{vanRitbergen:1998pn}.

The NNLO corrections are rather
small under these circumstances, amounting to less than 2\% for 
$N \geq 2$. At large $N$ the anomalous dimensions behave as
\beq
\label{eq:ntoinf}
  \gamma_{\,\rm ns}^{\,(n)\pm,\rm v}(N) \: = \:\: A_n (\ln N +\gamma_e) 
  - B_n - C_n \:\frac{\ln N}{N} + {\cal O} \left( \frac{1}{N} \right)
\eeq
where $\gamma_e$ is the Euler-Mascheroni constant and the coefficients 
are specified in the next paragraph.
Thus $\tilde{\gamma}_{\,\rm ns}^{\, +} = \gamma_{\,\rm ns}^{\, +} 
/ \ln N$, also shown in Fig.~1, approaches a constant for $N\ra\infty$.
The asymptotic results are indicated by replacing $\tilde{\gamma}
_{\,\rm ns}^{\,(n)+}(N=15)$ by $\tilde{\gamma}_{\,\rm ns}^{\,(n)+}
(N\ra\infty)$ for the respective highest term included in the 
curves (e.g., for $n=2$ at NNLO). Obviously the approach to the 
asymptotic limit is very slow. Yet the results at $N\ra\infty$, which 
can be derived much easier than the full $N$-dependence 
\cite{Berger:2002sv}, do provide a reasonable first estimate of the 
corrections.

The leading large-$N$ coefficients $A_n$, which are also relevant for 
the soft-gluon (threshold) resummation~\cite{Sterman:1987aj,%
Catani:1989ne,Catani:1991rp,Vogt:2000ci}, are given by
\bea
\label{eq:An}
  A_1 & = & 4\, C_F \nn \\[1.5mm]
  A_2 & = & 8\, C_F \left[ \left( \frac{67}{18} - \zeta_2^{} \right)
                C_A - \frac{5}{9}\,\nf \right] \nn \\[1.5mm]
  A_3 & = & 
     16\, C_F C_A^{\,2} \, \left( \frac{245}{24} - \frac{67}{9}\: \z2
         + \frac{11}{6}\:\z3 + \frac{11}{5}\:\z2^{\!\! 2} \right)
   \: + \: 16\, C_F^{\,2} \nf\, \left( -  \frac{55}{24}  + 2\:\z3 
   \right) \nn\\[1mm] & & \mbox{}
   + \: 16\, C_F C_A \nf\, \left( - \frac{209}{108} 
         + \frac{10}{9}\:\z2 - \frac{7}{3}\:\z3 \right)
   \: + \: 16\, C_F \n2f \left( - \frac{1}{27}\,\right) \:\: .
%
%
\eea
The $\nf$-independent contribution to the three-loop coefficient $A_3$ is 
also a new result of the present article. Inserting the numerical 
values of the $\zeta$-function and the QCD colour factors it reads 
$A_3|_{\nf=0}\cong 1174.898$, in agreement with the previous numerical estimate
of Ref.~\cite{Vogt:2000ci}. The constants $B_n$ can be read off directly
from the terms with $\delta(1-x)$ in Eqs.~(\ref{eq:Pqq0}), 
(\ref{eq:Pqq1p}) and (\ref{eq:Pqq2p}) below. Surprisingly, the 
coefficients $C_n$ in Eq.~(\ref{eq:ntoinf}), which are also best 
determined using those $x$-space results, turn out to be related to the 
$A_n$ by
\beq
\label{eq:Cn}
  C_1 \:\: = \:\: 0 \:\: , \quad
  C_2 \:\: = \:\: 4\, C_F\: A_1 \:\: , \quad
  C_3 \:\: = \:\: 8\, C_F\: A_2 \:\: .
\eeq
Especially the relation for $C_3$ is very suggestive and seems to call 
for a structural explanation. 
%
%
\setcounter{equation}{0}
\section{Results in {\bf x}-space}
\label{sec:xresults}
%
%
The splitting functions $P_{\,\rm ns}^{\,(n)\pm,s}(x)$ are obtained 
from the $N$-space results of the previous section by an inverse Mellin
transformation, which expresses these functions in terms of harmonic 
polylogarithms~\cite{Goncharov,Borwein,Remiddi:1999ew}.
The inverse Mellin transformation exploits an isomorphism between the 
set of harmonic sums for even or odd $N$ and the set of harmonic 
polylogarithms.  Hence it can be performed by a completely algebraic 
procedure~\cite{Moch:1999eb,Remiddi:1999ew}, based on the fact that 
harmonic sums occur as coefficients of the Taylor expansion of harmonic 
polylogarithms.

Our notation for the harmonic polylogarithms $H_{m_1,...,m_w}(x)$, 
$m_j = 0,\pm 1$ follows Ref.~\cite{Remiddi:1999ew} to which the reader 
is referred for a detailed discussion. 
The lowest-weight ($w = 1$) functions $H_m(x)$ are given by
\beq
\label{eq:hpol1}
  H_0(x)       \: = \: \ln x \:\: , \quad\quad
  H_{\pm 1}(x) \: = \: \mp \, \ln (1 \mp x) \:\: .
\eeq
The higher-weight ($w \geq 2$) functions are recursively defined as
\beq
\label{eq:hpol2}
  H_{m_1,...,m_w}(x) \: = \:
    \left\{ \begin{array}{cl}
    \displaystyle{ \frac{1}{w!}\,\ln^w x \:\: ,}
       & \quad {\rm if} \:\:\: m^{}_1,...,m^{}_w = 0,\ldots ,0 \\[2ex]
    \displaystyle{ \int_0^x \! dz\: f_{m_1}(z) \, H_{m_2,...,m_w}(z)
       \:\: , } & \quad {\rm else}
    \end{array} \right.
\eeq
with
\beq
\label{eq:hpolf}
  f_0(x)       \: = \: \frac{1}{x} \:\: , \quad\quad
  f_{\pm 1}(x) \: = \: \frac{1}{1 \mp x} \:\: .
\eeq
A useful short-hand notation is
\beq
\label{eq:habbr}
  H_{{\footnotesize \underbrace{0,\ldots ,0}_{\scriptstyle m} },\,
  \pm 1,\, {\footnotesize \underbrace{0,\ldots ,0}_{\scriptstyle n} },
  \, \pm 1,\, \ldots}(x) \: = \: H_{\pm (m+1),\,\pm (n+1),\, \ldots}(x)
  \:\: .
\eeq
For $w \leq 3$  the harmonic polylogarithms can be expressed in terms 
of standard polylogarithms; a complete list can be found in appendix A 
of Ref.~\cite{Moch:1999eb}. All harmonic polylogarithms of weight 
$w = 4$ in this article can be expressed in terms of standard 
polylogarithms, Nielsen functions~\cite{Lewin} or, by means of the 
defining relation (\ref{eq:hpol2}), as one-dimensional integrals
over these functions. 
A {\sc Fortran} program for the functions up to weight $w=4$ has been 
provided in Ref.~\cite{Gehrmann:2001pz}.

For completeness we recall the one- and two-loop non-singlet splitting
functions~\cite{Altarelli:1977zs,Curci:1980uw}
\bea
  P^{\,(0)}_{\,\rm ns}(x) & \! = \! &
  \colour4colour{\cf} \* \big( \,
          2 \* \pqq(x)
          + 3\* \delta(1 - x)
          \,\big)
\label{eq:Pqq0}
\eea
and 
\bea
\label{eq:Pqq1p}
  &&P^{\,(1)+}_{\,\rm ns}(x) \:\: = \:  
  4 \, \* \colour4colour{\ca \* \cf} \* \bigg(
         \pqq(x) \* \bigg[
            {67 \over 18}
          - \z2
          + {11 \over 6} \* \H(0)
          + \Hh(0,0)
          \bigg]
       + \pqq(- x) \* \bigg[
            \z2
          + 2 \* \Hh(-1,0)
          - \Hh(0,0)
          \bigg]
  \nonumber\\&& \mbox{} \quad
       + {14 \over 3} \* (1-x) 
       + \delta(1 - x) \* \bigg[
            {17 \over 24}
          + {11 \over 3} \* \z2
          - 3 \* \z3
          \bigg]
          \bigg)
- 4 \, \* \colour4colour{\cf \* \nf} \* \bigg(
         \pqq(x) \* \bigg[
            {5 \over 9}
          + {1 \over 3} \* \H(0)
          \bigg]
       + {2 \over 3} \* (1-x)
  \nonumber\\&& \mbox{} \quad
       + \delta(1 - x) \* \bigg[
            {1 \over 12}
          + {2 \over 3} \* \z2
          \bigg]
          \bigg)
+ 4 \, \* \colour4colour{\cf^{2}} \* \bigg(
         2 \* \pqq(x) \* \bigg[
            \Hh(1,0)
          - {3 \over 4} \* \H(0)
          + \H(2)
          \bigg]
       - 2 \* \pqq(-x) \* \bigg[
            \z2
          + 2 \* \Hh(-1,0)
  \quad \nonumber\\&& \mbox{} \quad
          - \Hh(0,0)
          \bigg]
       - (1-x) \* \bigg[
            1
          - {3 \over 2} \* \H(0)
          \bigg]
          - \H(0) 
          - (1+x) \* \Hh(0,0)
       + \delta(1 - x) \* \bigg[
           {3 \over 8}
          - 3 \* \z2
          + 6 \* \z3
          \bigg]
          \bigg)
 \:\: ,  \\[2mm] &&P^{\,(1)-}_{\,\rm ns}(x) \:\: = \:\:  
	P^{\,(1)+}_{\,\rm ns}(x)
       + 16\, \*  \colour4colour{\cf \* \bigg(\cf - {\ca \over 2} \bigg)}  \*  \bigg(
         \pqq(-x) \* \bigg[
            \z2
          + 2 \* \Hh(-1,0)
          - \Hh(0,0)
          \bigg]
          - 2 \* (1-x) 
  \nonumber\\&& \mbox{} \quad
          - (1+x) \* \H(0)
          \bigg)
\:\: .\label{eq:Pqq1m}
\eea
Here and in Eqs.~(\ref{eq:Pqq2p}) -- (\ref{eq:Pqq2s}) we suppress the
argument $x$ of the polylogarithms and use
\beq
  p_{\rm{qq}}(x) \: = \: 2\, (1 - x)^{-1} - 1 - x \:\: . 
\eeq
All divergences for $x \to 1 $ are understood in the sense of
$+$-distributions.

The three-loop splitting function for the evolution of the `plus'
combinations of quark densities in Eq.~(\ref{eq:qpm}), corresponding to
the anomalous dimension (\ref{eq:gqq2m}) reads
\bea
  &&P^{\,(2)+}_{\,\rm ns}(x) \:\: = \:\: 
         16 \, \* \colour4colour{\ca \* \cf \* \nf}  \*  \bigg(
           {1 \over 6} \* \pqq(x) \* \bigg[
            {10 \over 3} \* \z2
          - {209 \over 36}
          - 9 \* \z3
          - {167 \over 18} \* \H(0)
          + 2 \* \H(0) \* \z2
          - 7 \* \Hh(0,0)
          - 2 \* \Hhh(0,0,0)
  \nonumber\\&& \mbox{}
          + 3 \* \Hhh(1,0,0)
          - \H(3)
          \bigg]
         + {1 \over 3} \* \pqq(-x) \* \bigg[
            {3 \over 2} \* \z3
          - {5 \over 3} \* \z2
          - \Hh(-2,0)
          - 2 \* \H(-1) \* \z2
          - {10 \over 3} \* \Hh(-1,0)
          - \Hhh(-1,0,0)
  \nonumber\\&& \mbox{}
          + 2 \* \Hh(-1,2)
          + {1 \over 2} \* \H(0) \* \z2
          + {5 \over 3} \* \Hh(0,0)
          + \Hhh(0,0,0)
          - \H(3)
          \bigg]
       + (1-x) \* \bigg[
            {1 \over 6} \* \z2 
          - {257 \over 54}
          - {43 \over 18} \* \H(0)
          - {1 \over 6} \* \Hh(0,0)
          - \H(1)
          \bigg]
  \nonumber\\&& \mbox{}
       - (1+x) \* \bigg[
            {2 \over 3} \* \Hh(-1,0)
          + {1 \over 2} \* \H(2)
          \bigg]
          + {1 \over 3} \* \z2
	  + \H(0) 
          + {1 \over 6} \* \Hh(0,0) 
       + \delta(1 - x) \* \bigg[
            {5 \over 4}
          - {167 \over 54} \* \z2
          + {1 \over 20} \* \z2^2
          + {25 \over 18} \* \z3
          \bigg]
          \bigg)
  \nonumber\\&& \mbox{}
       + 16 \, \* \colour4colour{\ca \* \cf^2}  \*  \bigg(
           \pqq(x) \* \bigg[
            {5 \over 6} \* \z3
          - {69 \over 20} \* \z2^2
          - \Hh(-3,0)
          - 3 \* \H(-2) \* \z2
          - 14 \* \Hhh(-2,-1,0)
          + 3 \* \Hh(-2,0)
          + 5 \* \Hhh(-2,0,0)
  \nonumber\\&& \mbox{}
          - 4 \* \Hh(-2,2)
          - {151 \over 48} \* \H(0)
          + {41 \over 12} \* \H(0) \* \z2
          - {17 \over 2} \* \H(0) \* \z3
          - {13 \over 4} \* \Hh(0,0)
          - 4 \* \Hh(0,0) \* \z2
          - {23 \over 12} \* \Hhh(0,0,0)
          + 5 \* \Hhhh(0,0,0,0)
          + {2 \over 3} \* \H(3)
  \nonumber\\&& \mbox{}
          - 24 \* \H(1) \* \z3
          - 16 \* \Hhh(1,-2,0)
          + {67 \over 9} \* \Hh(1,0)
          - 2 \* \Hh(1,0) \* \z2
          + {31 \over 3} \* \Hhh(1,0,0)
          + 11 \* \Hhhh(1,0,0,0)
          + 8 \* \Hhhh(1,1,0,0)
          - 8 \* \Hh(1,3)
          + \H(4)
  \nonumber\\&& \mbox{}
          + {67 \over 9} \* \H(2)
          - 2 \* \H(2) \* \z2
          + {11 \over 3} \* \Hh(2,0)
          + 5 \* \Hhh(2,0,0)
          + \Hh(3,0)
          \bigg]
         + \pqq(-x) \* \bigg[
            {1 \over 4} \* \z2^2
          - {67 \over 9} \* \z2
          + {31 \over 4} \* \z3
          + 5 \* \Hh(-3,0)
  \nonumber\\&& \mbox{}
          - 32 \* \H(-2) \* \z2
          - 4 \* \Hhh(-2,-1,0)
          - {31 \over 6} \* \Hh(-2,0)
          + 21 \* \Hhh(-2,0,0)
          + 30 \* \Hh(-2,2)
          - {31 \over 3} \* \H(-1) \* \z2
          - 42 \* \H(-1) \* \z3
          + {9 \over 4} \* \H(0)
  \nonumber\\&& \mbox{}
          - 4 \* \Hhh(-1,-2,0)
          + 56 \* \Hh(-1,-1) \* \z2
          - 36 \* \Hhhh(-1,-1,0,0)
          - 56 \* \Hhh(-1,-1,2)
          - {134 \over 9} \* \Hh(-1,0)
          - 42 \* \Hh(-1,0) \* \z2
          - \Hh(3,0)
  \nonumber\\&& \mbox{}
          + 32 \* \Hh(-1,3)
          - {31 \over 6} \* \Hhh(-1,0,0)
          + 17 \* \Hhhh(-1,0,0,0)
          + {31 \over 3} \* \Hh(-1,2)
          + 2 \* \Hhh(-1,2,0)
          + {13 \over 12} \* \H(0) \* \z2
          + {29 \over 2} \* \H(0) \* \z3
          + {67 \over 9} \* \Hh(0,0)
  \nonumber\\&& \mbox{}
          + 13 \* \Hh(0,0) \* \z2
          + {89 \over 12} \* \Hhh(0,0,0)
          - 5 \* \Hhhh(0,0,0,0)
          - 7 \* \H(2) \* \z2
          - {31 \over 6} \* \H(3)
          - 10 \* \H(4)
          \bigg]
       + (1-x) \* \bigg[
            {133 \over 36}
          + 4 \* \Hhhh(0,0,0,0)
  \nonumber\\&& \mbox{}
          - {167 \over 4} \* \z3 
          - 2 \* \H(0) \* \z3
          - 2 \* \Hh(-3,0)
          + \H(-2) \* \z2
          + 2 \* \Hhh(-2,-1,0)
          - 3 \* \Hhh(-2,0,0)
          + {77 \over 4} \* \Hhh(0,0,0) 
          - {209 \over 6} \* \H(1)
          - 7 \* \H(1) \* \z2
  \nonumber\\&& \mbox{}
          + 4 \* \Hhh(1,0,0)
          + {14 \over 3} \* \Hh(1,0)
          \bigg]
       + (1+x) \* \bigg[
            {43 \over 2} \* \z2
          - 3 \* \z2^2
          + {25 \over 2} \* \Hh(-2,0)
          - 31 \* \H(-1) \* \z2
          - 14 \* \Hhh(-1,-1,0)
          - {13 \over 3} \* \Hh(-1,0)
  \nonumber\\&& \mbox{}
          + 24 \* \Hh(-1,2)
          + 23 \* \Hhh(-1,0,0)
          + {55 \over 2} \* \H(0) \* \z2 
          + 5 \* \Hh(0,0) \* \z2
          + {1457 \over 48} \* \H(0)
          - {1025 \over 36} \* \Hh(0,0)
          - {155 \over 6} \* \H(2) 
          + \H(2) \* \z2
          - 15 \* \H(3)
  \nonumber\\&& \mbox{}
          + 2 \* \Hhh(2,0,0)
          - 3 \* \H(4)
          \bigg]
          - 5 \* \z2
          - {1 \over 2} \* \z2^2
          + 50 \* \z3
          - 2 \* \Hh(-3,0)
          - 7 \* \Hh(-2,0)
          - \H(0) \* \z3
          - {37 \over 2} \* \H(0) \* \z2 
          - {242 \over 9} \* \H(0)
  \nonumber\\&& \mbox{}
          - 2 \* \Hh(0,0) \* \z2
          + {185 \over 6} \* \Hh(0,0)
          - 22 \* \Hhh(0,0,0)
          - 4 \* \Hhhh(0,0,0,0)
          + {28 \over 3} \* \H(2)
          + 6 \* \H(3)
       + \delta(1 - x) \* \bigg[
            {151 \over 64}
          + \z2 \* \z3
          - {205 \over 24} \* \z2
  \nonumber\\&& \mbox{}
          - {247 \over 60} \* \z2^2
          + {211 \over 12} \* \z3
          + {15 \over 2} \* \z5
          \bigg]
          \bigg)
       + 16 \, \* \colour4colour{\ca^2 \* \cf}  \*  \bigg(
           \pqq(x) \* \bigg[
            {245 \over 48}
          - {67 \over 18} \* \z2
          + {12 \over 5} \* \z2^2
          + {1 \over 2} \* \z3
          + {1043 \over 216} \* \H(0)
  \nonumber\\&& \mbox{}
          + \Hh(-3,0)
          + 4 \* \Hhh(-2,-1,0)
          - {3 \over 2} \* \Hh(-2,0)
          - \Hhh(-2,0,0)
          + 2 \* \Hh(-2,2)
          - {31 \over 12} \* \H(0) \* \z2
          + 4 \* \H(0) \* \z3
          + {389 \over 72} \* \Hh(0,0)
          - 2 \* \Hhh(2,0,0)
  \nonumber\\&& \mbox{}
          - \Hhhh(0,0,0,0)
          + 9 \* \H(1) \* \z3
          + 6 \* \Hhh(1,-2,0)
          - \Hh(1,0) \* \z2
          - {11 \over 4} \* \Hhh(1,0,0)
          - 3 \* \Hhhh(1,0,0,0)
          - 4 \* \Hhhh(1,1,0,0)
          + 4 \* \Hh(1,3)
          + {31 \over 12} \* \Hhh(0,0,0)
  \nonumber\\&& \mbox{}
          + {11 \over 12} \* \H(3)
          + \H(4)
          \bigg]
         + \pqq(-x) \*  \bigg[
            {67 \over 18} \*  \z2
          - \z2^2
          - {11 \over 4} \* \z3
          - \Hh(-3,0)
          + 8 \* \H(-2) \* \z2
          + {11 \over 6} \* \Hh(-2,0)
          - 4 \* \Hhh(-2,0,0)
  \nonumber\\&& \mbox{}
          - 3 \* \Hhhh(-1,0,0,0)
          + {11 \over 3} \* \H(-1) \* \z2
          + 12 \* \H(-1) \* \z3
          - 16 \* \Hh(-1,-1) \* \z2
          + 8 \* \Hhhh(-1,-1,0,0)
          + 16 \* \Hhh(-1,-1,2)
          + {67 \over 9} \* \Hh(-1,0)
  \nonumber\\&& \mbox{}
          - 8 \* \Hh(-2,2)
          + 11 \* \Hh(-1,0) \* \z2
          + {11 \over 6} \* \Hhh(-1,0,0)
          - {11 \over 3} \* \Hh(-1,2)
          - 8 \* \Hh(-1,3)
          - {3 \over 4} \* \H(0)
          - {1 \over 6} \* \H(0) \* \z2
          - 4 \* \H(0) \* \z3
          - {67 \over 18} \* \Hh(0,0)
  \nonumber\\&& \mbox{}
          - 3 \* \Hh(0,0) \* \z2
          - {31 \over 12} \* \Hhh(0,0,0)
          + \Hhhh(0,0,0,0)
          + 2 \* \H(2) \* \z2
          + {11 \over 6} \* \H(3)
          + 2 \* \H(4)
          \bigg]
       + (1-x) \* \bigg[
            {1883 \over 108}
          - {1 \over 2} \* \Hhhh(0,0,0,0)
          + 11 \* \H(1) 
  \nonumber\\&& \mbox{}
          - \Hhh(-2,-1,0)
          + {1 \over 2} \* \Hh(-3,0)
          - {1 \over 2} \* \H(-2) \* \z2
          + {1 \over 2} \* \Hhh(-2,0,0)
          + {523 \over 36} \* \H(0) 
          + \H(0) \* \z3 
          - {13 \over 3} \* \Hh(0,0) 
          - {5 \over 2} \* \Hhh(0,0,0) 
          + 2 \* \H(1) \* \z2
  \nonumber\\&& \mbox{}
          - 2 \* \Hhh(1,0,0)
          \bigg]
       + (1+x) \* \bigg[
            8 \* \H(-1) \* \z2
          + 4 \* \Hhh(-1,-1,0)
          + {8 \over 3} \* \Hh(-1,0)
          - 5 \* \Hhh(-1,0,0)
          - 6 \* \Hh(-1,2)
          - {13 \over 3} \* \z2
          + {3 \over 8} \* \z2^2
  \nonumber\\&& \mbox{}
          - {43 \over 4} \* \z3
          - {5 \over 2} \* \Hh(-2,0)
          - {11 \over 2} \* \H(0) \* \z2 
          - {1 \over 2} \* \H(2) \* \z2
          - {5 \over 4} \* \Hh(0,0) \* \z2
          + 7 \* \H(2)
          - {1 \over 4} \* \Hhh(2,0,0)
          + 3 \* \H(3)
          + {3 \over 4} \* \H(4)
          \bigg]
          + {1 \over 2} \* \Hh(0,0) \* \z2 
  \nonumber\\&& \mbox{}
          + {1 \over 4} \* \z2^2
          - {8 \over 3} \* \z2
          + {17 \over 2} \* \z3
          + \Hh(-2,0)
          - {19 \over 2} \* \H(0)
          + {5 \over 2} \* \H(0) \* \z2
          - \H(0) \* \z3 
          + {13 \over 3} \* \Hh(0,0)
          + {5 \over 2} \* \Hhh(0,0,0)
          + {1 \over 2} \* \Hhhh(0,0,0,0)
  \nonumber\\&& \mbox{}
       - \delta(1 - x) \*  \bigg[
            {1657 \over 576}
          - {281 \over 27} \* \z2
          + {1 \over 8} \* \z2^2
          + {97 \over 9} \* \z3
          - {5 \over 2} \* \z5
          \bigg]
          \bigg)
       + 16 \, \* \colour4colour{\cf \* \n2f}  \*  \bigg(
            {1 \over 18} \* \pqq(x) \*  \bigg[
            \Hh(0,0)
          - {1 \over 3}
          + {5 \over 3} \* \H(0)
          \bigg]
  \nonumber\\&& \mbox{}
       + (1-x) \* \bigg[
            {13 \over 54}
          + {1 \over 9} \* \H(0)
          \bigg]
       - \delta(1 - x) \*  \bigg[
            {17 \over 144}
          - {5 \over 27} \* \z2
          + {1 \over 9} \* \z3
          \bigg]
          \bigg)
       + 16 \, \* \colour4colour{\cf^2 \* \nf}  \*  \bigg(
           {1 \over 3} \* \pqq(x) \*  \bigg[
            5 \* \z3
          - 4 \* \Hhh(1,0,0)
  \nonumber\\&& \mbox{}
          - {55 \over 16}
          + {5 \over 8} \* \H(0)
          + \H(0) \* \z2
          + {3 \over 2} \* \Hh(0,0)
          - \Hhh(0,0,0)
          - {10 \over 3} \* \Hh(1,0)
          - {10 \over 3} \* \H(2)
          - 2 \* \Hh(2,0)
          - 2 \* \H(3)
          \bigg]
         + {2 \over 3} \* \pqq(-x) \*  \bigg[
            {5 \over 3} \* \z2
  \nonumber\\&& \mbox{}
          - {3 \over 2} \* \z3
          + \Hh(-2,0)
          + 2 \* \H(-1) \* \z2
          + {10 \over 3} \* \Hh(-1,0)
          + \Hhh(-1,0,0)
          - 2 \* \Hh(-1,2)
          - {1 \over 2} \* \H(0) \* \z2
          - {5 \over 3} \* \Hh(0,0)
          - \Hhh(0,0,0)
          + \H(3)
          \bigg]
  \nonumber\\&& \mbox{}
       - (1-x) \* \bigg[
            {10 \over 9}
          + {19 \over 18} \* \Hh(0,0)
          - {4 \over 3} \* \H(1)
          + {2 \over 3} \* \Hh(1,0)
          + {4 \over 3} \* \H(2)
          \bigg]
       + (1+x) \* \bigg[
            {4 \over 3} \* \Hh(-1,0)
          - {25 \over 24} \* \H(0)
          + {1 \over 2} \* \Hhh(0,0,0)
          \bigg]
          + {2 \over 9} \* \H(0)
  \nonumber\\&& \mbox{} 
          + {7 \over 9} \* \Hh(0,0)
          + {4 \over 3} \* \H(2)
       - \delta(1 - x) \*  \bigg[
            {23 \over 16}
          - {5 \over 12} \* \z2
          - {29 \over 30} \* \z2^2
          + {17 \over 6} \* \z3
          \bigg]
          \bigg)
       + 16 \, \* \colour4colour{\cf^3}  \*  \bigg(
           \pqq(x) \*  \bigg[
            {9 \over 10} \* \z2^2
          - 2 \* \Hh(-3,0)
  \nonumber\\&& \mbox{}
          + 6 \* \H(-2) \* \z2
          + 12 \* \Hhh(-2,-1,0)
          - 6 \* \Hhh(-2,0,0)
          - {3 \over 16} \* \H(0)
          - {3 \over 2} \* \H(0) \* \z2
          + \H(0) \* \z3
          + {13 \over 8} \* \Hh(0,0)
          - 2 \* \Hhhh(0,0,0,0)
          + 8 \* \Hh(1,3)
  \nonumber\\ && \mbox{}
          + 12 \* \H(1) \* \z3
          + 8 \* \Hhh(1,-2,0)
          - 6 \* \Hhh(1,0,0)
          - 4 \* \Hhhh(1,0,0,0)
          + 4 \* \Hhh(1,2,0)
          - 3 \* \Hh(2,0)
          + 2 \* \Hhh(2,0,0)
          + 4 \* \Hhh(2,1,0)
          + 4 \* \Hh(2,2)
  \nonumber\\&& \mbox{}
          + 4 \* \Hh(3,0)
          + 4 \* \Hh(3,1)
          + 2 \* \H(4)
          \bigg]
         + \pqq(-x) \*  \bigg[
            {7 \over 2} \* \z2^2
          - {9 \over 2} \* \z3
          - 6 \* \Hh(-3,0)
          + 32 \* \H(-2) \* \z2
          + 8 \* \Hhh(-2,-1,0)
          + 3 \* \Hh(-2,0)
  \nonumber\\ && \mbox{}
          - 26 \* \Hhh(-2,0,0)
          - 28 \* \Hh(-2,2)
          + 6 \* \H(-1) \* \z2
          + 36 \* \H(-1) \* \z3
          + 8 \* \Hhh(-1,-2,0)
          - 48 \* \Hh(-1,-1) \* \z2
          + 40 \* \Hhhh(-1,-1,0,0)
  \nonumber\\&& \mbox{}
          + 48 \* \Hhh(-1,-1,2)
          + 40 \* \Hh(-1,0) \* \z2
          + 3 \* \Hhh(-1,0,0)
          - 22 \* \Hhhh(-1,0,0,0)
          - 6 \* \Hh(-1,2)
          - 4 \* \Hhh(-1,2,0)
          - 32 \* \Hh(-1,3)
          - {3 \over 2} \* \H(0)
  \nonumber\\&& \mbox{}
          - {3 \over 2} \* \H(0) \* \z2
          - 13 \* \H(0) \* \z3
          - 14 \* \Hh(0,0) \* \z2
          - {9 \over 2} \* \Hhh(0,0,0)
          + 6 \* \Hhhh(0,0,0,0)
          + 6 \* \H(2) \* \z2
          + 3 \* \H(3)
          + 2 \* \Hh(3,0)
          + 12 \* \H(4)
          \bigg]
  \nonumber\\&& \mbox{}
       + (1-x) \* \bigg[
            2 \* \Hh(-3,0)  
          - {31 \over 8}
          + 4 \* \Hhh(-2,0,0)
          + \Hh(0,0) \* \z2
          - 3 \* \Hhhh(0,0,0,0)
          + 35 \* \H(1)
          + 6 \* \H(1) \* \z2
          - \Hh(1,0)
          + {5 \over 2} \* \Hh(2,0)
          \bigg]
  \nonumber\\&& \mbox{}
       + (1+x) \* \bigg[
            {37 \over 10} \* \z2^2
          - {93 \over 4} \* \z2
          - {81 \over 2} \* \z3 
          - 15 \* \Hh(-2,0)
          + 30 \* \H(-1) \* \z2
          + 12 \* \Hhh(-1,-1,0)
          - 2 \* \Hh(-1,0)
          - 26 \* \Hhh(-1,0,0)
  \nonumber\\&& \mbox{}
          - 24 \* \Hh(-1,2)
          - {539 \over 16} \* \H(0)
          - 28 \* \H(0) \* \z2
          + {191 \over 8} \* \Hh(0,0)
          + 20 \* \Hhh(0,0,0)
          + {85 \over 4} \* \H(2)
          - 3 \* \Hhh(2,0,0)
          - 2 \* \Hh(3,0)
          + 13 \* \H(3)
  \nonumber\\&& \mbox{}
          - \H(4)
          \bigg]
          + 4 \* \z2
          + 33 \* \z3
          + 4 \* \Hh(-3,0)
          + 10 \* \Hh(-2,0)
          + {67 \over 2} \* \H(0)
          + 6 \* \H(0) \* \z3
          + 19 \* \H(0) \* \z2
          - 25 \* \Hh(0,0)
          - 17 \* \Hhh(0,0,0)
  \nonumber\\&& \mbox{}
          - 2 \* \H(2)
          - \Hh(2,0)
          - 4 \* \H(3)
       + \delta(1 - x) \*  \bigg[
            {29 \over 32}
          - 2 \* \z2 \* \z3
          + {9 \over 8} \* \z2
          + {18 \over 5} \* \z2^2
          + {17 \over 4} \* \z3
          - 15 \* \z5
          \bigg]
          \bigg)
\: \: .\label{eq:Pqq2p}
\eea
The $x$-space counterpart of Eq.~(\ref{eq:gqq2m}) for the evolution
of the `minus' combinations (\ref{eq:qpm}) is given~by
\bea
  &&P^{\,(2)-}_{\,\rm ns}(x) \:\: = \:\:  
	P^{\,(2)+}_{\,\rm ns}(x)
 + 16 \, \* \colour4colour{\ca \* \cf \* \bigg(\cf - {\ca \over 2} \bigg)}  \*  \bigg(
           \pqq(-x) \* \bigg[
            {134 \over 9} \* \z2
          - 4 \* \z2^2
          - 11 \* \z3
          - 4 \* \Hh(-3,0)
  \nonumber\\&& \mbox{}
          + 32 \* \H(-2) \* \z2
          + {22 \over 3} \* \Hh(-2,0)
          - 16 \* \Hhh(-2,0,0)
          - 32 \* \Hh(-2,2)
          + {44 \over 3} \* \H(-1) \* \z2
          + 48 \* \H(-1) \* \z3
          - 64 \* \Hh(-1,-1) \* \z2
  \nonumber\\&& \mbox{}
          + 32 \* \Hhhh(-1,-1,0,0)
          + 64 \* \Hhh(-1,-1,2)
          + {268 \over 9} \* \Hh(-1,0)
          + 44 \* \Hh(-1,0) \* \z2
          + {22 \over 3} \* \Hhh(-1,0,0)
          - 12 \* \Hhhh(-1,0,0,0)
          - {44 \over 3} \* \Hh(-1,2)
  \nonumber\\&& \mbox{}
          - 32 \* \Hh(-1,3)
          - 3 \* \H(0)
          - {2 \over 3} \* \H(0) \* \z2
          - 16 \* \H(0) \* \z3
          - {134 \over 9} \* \Hh(0,0)
          - 12 \* \Hh(0,0) \* \z2
          - {31 \over 3} \* \Hhh(0,0,0)
          + 4 \* \Hhhh(0,0,0,0)
          + 8 \* \H(2) \* \z2
  \nonumber\\&& \mbox{}
          + {22 \over 3} \* \H(3)
          + 8 \* \H(4)
          \bigg]
         + (1-x) \* \bigg[
            {367 \over 18}
          + {1 \over 2} \* \z2^2
          + 2 \* \Hh(-3,0)
          - 2 \* \H(-2) \* \z2
          - 4 \* \Hhh(-2,-1,0)
          - 10 \* \Hh(-2,0)
          - 2 \* \Hh(0,0)
  \nonumber\\&& \mbox{}
          + 2 \* \Hhh(-2,0,0)
          + 2 \* \H(0) \* \z3
          + \Hh(0,0) \* \z2
          - \Hhhh(0,0,0,0)
          + 8 \* \H(1) \* \z2
          + {140 \over 3} \* \H(1)
          \bigg]
         + (1+x) \* \bigg[
            32 \* \H(-1) \* \z2
          - 18 \* \z2
  \nonumber\\&& \mbox{}
          - 23 \* \z3
          + {26 \over 3} \* \Hh(-1,0)
          - 16 \* \Hhh(-1,0,0)
          - 32 \* \Hh(-1,2)
          - {481 \over 18} \* \H(0)
          - 29 \* \H(0) \* \z2
          + 5 \* \Hhh(0,0,0)
          + 24 \* \H(3)
          + {70 \over 3} \* \H(2)
          \bigg]
  \nonumber\\&& \mbox{}
          - 2 \* \z2
          - 2 \* \z3
          + 32 \* \H(0)
          + 14 \* \H(0) \* \z2
          + 2 \* \Hhh(0,0,0)
          - 16 \* \H(3)
          \bigg)
  + 16 \, \* \colour4colour{\cf \* \nf \* \bigg(\cf - {\ca \over 2} \bigg)}  \*  \bigg(
           \pqq(-x) \* \bigg[
            2 \* \z3
  \nonumber\\&& \mbox{}
          - {20 \over 9} \* \z2
          - {4 \over 3} \* \Hh(-2,0)
          - {8 \over 3} \* \H(-1) \* \z2
          - {40 \over 9} \* \Hh(-1,0)
          - {4 \over 3} \* \Hhh(-1,0,0)
          + {8 \over 3} \* \Hh(-1,2)
          + {2 \over 3} \* \H(0) \* \z2
          + {20 \over 9} \* \Hh(0,0)
          + {4 \over 3} \* \Hhh(0,0,0)
  \nonumber\\&& \mbox{}
          - {4 \over 3} \* \H(3)
	  \bigg]
           + (1-x) \* \bigg[
            {61 \over 9}
          - {8 \over 3} \* \H(1)
	  \bigg]
           + (1+x) \* \bigg[
            2 \* \Hh(0,0)
          - {8 \over 3} \* \Hh(-1,0)
          + {41 \over 9} \* \H(0)
          - {4 \over 3} \* \H(2)
	  \bigg]
          \bigg)
  \nonumber\\&& \mbox{}
  + 16 \, \* \colour4colour{\cf^2 \* \bigg(\cf - {\ca \over 2} \bigg)}  \*  \bigg(
           \pqq(-x) \* \bigg[
            9 \* \z3
          - 7 \* \z2^2
          + 12 \* \Hh(-3,0)
          - 64 \* \H(-2) \* \z2
          - 16 \* \Hhh(-2,-1,0)
          - 6 \* \Hh(-2,0)
  \nonumber\\ && \mbox{}
          + 52 \* \Hhh(-2,0,0)
          + 56 \* \Hh(-2,2)
          - 12 \* \H(-1) \* \z2
          - 72 \* \H(-1) \* \z3
          - 16 \* \Hhh(-1,-2,0)
          + 96 \* \Hh(-1,-1) \* \z2
          - 80 \* \Hhhh(-1,-1,0,0)
  \nonumber\\ && \mbox{}
          - 96 \* \Hhh(-1,-1,2)
          - 80 \* \Hh(-1,0) \* \z2
          - 6 \* \Hhh(-1,0,0)
          + 44 \* \Hhhh(-1,0,0,0)
          + 12 \* \Hh(-1,2)
          + 8 \* \Hhh(-1,2,0)
          + 64 \* \Hh(-1,3)
          + 3 \* \H(0)
  \nonumber\\ && \mbox{}
          + 3 \* \H(0) \* \z2
          + 26 \* \H(0) \* \z3
          + 28 \* \Hh(0,0) \* \z2
          + 9 \* \Hhh(0,0,0)
          - 12 \* \Hhhh(0,0,0,0)
          - 12 \* \H(2) \* \z2
          - 6 \* \H(3)
          - 4 \* \Hh(3,0)
          - 24 \* \H(4)
	  \bigg]
  \nonumber\\&& \mbox{}
           - (1-x) \* \bigg[
            15
          + 8 \* \Hh(-3,0)
          + 8 \* \Hhh(-2,0,0)
          + 61 \* \H(0)
          + 6 \* \H(0) \* \z3
          + 2 \* \Hh(0,0) \* \z2
          - 6 \* \Hhhh(0,0,0,0)
          + 12 \* \H(1) \* \z2
          + 60 \* \H(1)
  \nonumber\\&& \mbox{}
          + 8 \* \Hh(1,0)
	  \bigg]
           + (1+x) \* \bigg[
            24 \* \z2 
          + 57 \* \z3
          + 10 \* \Hh(-2,0)
          - 48 \* \H(-1) \* \z2
          - 4 \* \Hh(-1,0)
          + 40 \* \Hhh(-1,0,0)
          + 48 \* \Hh(-1,2)
  \nonumber\\&& \mbox{}
          + 59 \* \H(0) \* \z2
          - 22 \* \Hh(0,0)
          - 35 \* \Hhh(0,0,0)
          - 22 \* \H(2)
          - 4 \* \Hh(2,0)
          - 44 \* \H(3)
	  \bigg]
          + 8 \* \z2
          - 42 \* \z3
          - 4 \* \H(-2,0)
          + 42 \* \H(0)
  \nonumber\\&& \mbox{}
          - 38 \* \H(0) \* \z2
          + 14 \* \Hh(0,0)
          - 16 \* \H(2)
          + 26 \* \Hhh(0,0,0)
          + 24 \* \H(3)
          \bigg)
 \:\: .\label{eq:Pqq2m}
\eea
Finally the Mellin inversion of $\gamma^{(2)\rm s}_{\,\rm ns}(N)$ in
Eq.~(\ref{eq:gqq2s}) leads to the following result for the leading
(third-order) difference $P^{\,(2){\rm s}}_{\,\rm ns}(x)$ of the 
`valence' and `minus' splitting functions:
\bea
  &&P^{\,(2){\rm s}}_{\,\rm ns}(x) \:\: = \:\:  
         16\, \*  \colour4colour{\nf \, \* \dabc2n}  \*  \bigg(
         {1 \over 2} \* (1-x) \* \bigg[
            {50 \over 3}
          + {41 \over 12} \* \z2
          - {5 \over 4} \* \z2^2 
          - \Hh(-3,0)
          + \H(-2) \* \z2
          - \Hhh(-2,0,0)
          + {9 \over 4} \* \H(3) 
  \nonumber\\&& \mbox{}
          + 2 \* \Hhh(-2,-1,0)
          + {3 \over 2} \* \Hh(0,0) \* \z2
          - {1 \over 2} \* \H(1) \* \z2
          - {3 \over 4} \* \Hhh(1,0,0)
          + {91 \over 12} \* \H(1)
          \bigg]
       + {1 \over 2} \* (1+x) \* \bigg[
            \Hhh(-1,-1,0)
          - {3 \over 2} \* \H(-1) \* \z2
          + {3 \over 4} \* \H(0)
  \nonumber\\&& \mbox{}
          - {13 \over 6} \* \Hh(-1,0)
          + {1 \over 2} \* \Hhh(-1,0,0)
          + 2 \* \Hh(-1,2)
          - {3 \over 2} \* \Hh(-2,0)
          + {9 \over 4} \* \H(0) \* \z2
          + {29 \over 12} \* \Hh(0,0)
          + {41 \over 12} \* \H(2)
          - \H(2) \* \z2
          - {1 \over 2} \* \Hhh(2,0,0)
  \nonumber\\&& \mbox{}
          + {3 \over 2} \* \H(4)
          \bigg]
       - {1 \over 3} \* \bigg( {1 \over x}+x^2 \bigg) \* \bigg[
            3 \* \H(-1) \* \z2
          + 2 \* \Hhh(-1,-1,0)
          - 2 \* \Hhh(-1,0,0)
          - 2 \* \Hh(-1,2)
          + \H(1) \* \z2
          \bigg]
       + {1 \over 3} \* x^2 \* \bigg[
            5 \* \z3
          - 2 \* \H(3)
  \nonumber\\&& \mbox{}
          + 2 \* \Hh(-2,0)
          + 4 \* \H(0) \* \z2
          - 2 \* \Hhh(0,0,0)
          + 2 \* \H(1) \* \z2
          \bigg]
          + {91 \over 24} \* \H(0)
          + \z3
          - {9 \over 2} \* \z2
          + \z2^2
          - \H(0) \* \z3
          - \H(0) \* \z2
          - 2 \* \Hh(0,0) \* \z2
  \nonumber\\&& \mbox{}
          + {3 \over 8} \* \Hh(0,0)
          - {1 \over 4} \* \Hhh(0,0,0)
          + {1 \over 2} \* \Hhhh(0,0,0,0)
          + \Hh(-2,0)
          - \H(3)
          \bigg)
 \:\: . \label{eq:Pqq2s}
\eea

Of particular interest is the end-point behaviour of the harmonic polylogarithms
at $x\to 0$ or $x\to 1$, where logarithmic singularities occur.  In the limit 
$x\to 0$, the factors $\ln x$ are related to trailing zeroes in the index 
field, whereas in the limit $x \to 1$ factors of $\ln(1-x)$ emerge from leading 
indices of value 1.  In both limits, the logarithms can be factored out by 
repeated use of the product identity for harmonic polylogarithms, 
\bea
 H_{\vec{m}_w}(x)H_{\vec{n}_v}(x) &\! = \! & \sum_{{\vec{l}_{w+v}}\, =\, 
 \vec{m}_w \uplus \vec{n}_v} H_{\vec{l}_{w+v}}(x) \:\: .
\label{eq:halgebra} 
\eea
Here $\vec{m}_w \uplus \vec{n}_v$ represents all mergers of $\vec{m}_w$ 
and $\vec{n}_v$ in which the relative orders of the elements of $\vec{m}_w$ 
and $\vec{n}_v$ are preserved. All algorithms for this algebraic procedure 
have been coded in {\sc Form}, some explicit examples are given 
in Refs.~\cite{Remiddi:1999ew,Blumlein:2003gb}.

The large-$x$ behaviour of splitting functions $P^{(n)\pm,\rm v}_{\,\rm ns}(x)$ 
reflects the large-$N$ behaviour of the corresponding anomalous dimensions in 
Eq.~(\ref{eq:ntoinf}).
Specifically, the (identical) large-$x$ behaviour of 
$P^{(2)\pm,\rm v}_{\,\rm ns}(x)$ is given by
\beq
\label{eq:Plargex}
  P^{(2)\pm,\rm v}_{x\ra 1}(x) \: = \: 
     \frac{A_3}{(1-x)_+} \: + \:  B_3\: \delta(1-x) \: 
     + \: C_3\: \ln (1-x) \: + \: {\cal O}(1) \:\: .
\eeq
The constants $A_3$ and $C_3$ have been specified in Eqs.~(\ref{eq:An})
and (\ref{eq:Cn}), respectively, while the coefficients of $\delta(1-x)$
are explicit in Eq.~(\ref{eq:Pqq2p}). At small $x$ the splitting functions 
can be expanded in powers of $\ln x$. For the three-loop non-singlet
splitting functions $P^{(n)\pm,\rm s}_{\,\rm ns}(x)$ one finds
\beq
\label{eq:Plnx}
  P^{\,(2)i}_{x\ra 0}(x) \: = \: D_0^{\, i} \ln^4 x \: 
  + \: D_1^{\, i} \ln^3 x \: + \: D_2^{\, i} \ln^2 x \: 
  + \: D_3^{\, i} \ln x \: + \: {\cal O}(1) \:\: .
\eeq
Generally, terms up to $\ln^{\, 2k} x$ occur at order 
$\alpha_{\rm s}^{\, k+1}$. Keeping only the highest $n\! +\! 1$ of 
these, one arrives at the N$^{\rm n}$Lx small-$x$ approximation. Like 
the large-$x$ coefficients, these contributions can be readily extracted
from our full results using Eq.~(\ref{eq:halgebra}). 
For $P^{\,(2)+}_{\,\rm ns}$ we~obtain
\bea
\label{eq:D+exact}
  D_0^+ & = &   \frac{2}{3}\: C_F^{\, 3}  \nonumber \\[1.5mm]
  D_1^+ & = &   \frac{22}{3}\: C_F^{\, 2} \ca - 4\: C_F^{\, 3} 
              - \frac{4}{3}\: C_F^{\, 2} \nf  \nonumber \\[1.5mm]
  D_2^+ & = &   \left[ \frac{121}{9} - 30\,\z2 \right] \cf C_A^{\, 2}
              + \left[ \frac{472}{9} + 96\,\z2 \right] C_F^{\, 2} \ca
              + [ 4 - 104\, \z2 ]\: C_F^{\, 3} 
              - \frac{44}{9}\, \cf \ca \nf
 \nonumber \\ & & \mbox{}
              - \frac{64}{9}\, C_F^{\, 2} \nf 
              + \frac{4}{9} \cf n_{\! f}^{\, 2}  \nonumber \\[1.5mm]
  D_3^+ & = &   \left[ \frac{3934}{27} - 92\,\z2 \right] \cf C_A^{\, 2}
              + \left[ \frac{370}{9} + 216\,\z2 + 48\,\z3 \right]
                C_F^{\, 2} \ca
 \\ & & \mbox{}
              - [ 30 + 192\,\z2 + 96\,\z3 ]\: C_F^{\, 3}
              - \left[ \frac{1268}{27} - 8\,\z2 \right] \cf \ca \nf
              - \frac{88}{9}\, C_F^{\, 2} \nf 
              + \frac{88}{27}\, \cf n_{\! f}^{\, 2} 
 \:\: , \:\:\:\: \nonumber
\eea
or, after inserting $C_A=3$ and $C_F=4/3$ and the numerical values
of $\z2$ and $\z3$,
\bea
\label{eq:D+num}
  D_0^+ &\cong & 1.58025  \nonumber \\
  D_1^+ &\cong & 29.6296 - 2.37037\: \nf  \nonumber \\
  D_2^+ &\cong & 295.042 - 32.1975\: \nf + 0.592592\: n_{\! f}^{\, 2}
  \nonumber \\
  D_3^+ &\cong & 1261.11 - 152.597\: \nf + 4.345679\: n_{\! f}^{\, 2}
  \:\: .
\eea
The corresponding coefficients for $P^{\,(2)-}_{\,\rm ns}$ are given by
\bea
\label{eq:D-exact}
  D_0^- & = & \mbox{} - \cf C_A^{\, 2} + 4\: C_F^{\, 2} \ca 
              - \frac{10}{3}\: C_F^{\, 3}   \nonumber \\[1.5mm]
  D_1^- & = &   \frac{40}{9}\: \cf C_A^{\, 2} 
              - \frac{14}{9}\: C_F^{\, 2} \ca - 4\: C_F^{\, 3} 
              + \frac{20}{9}\: C_F^{\, 2} \nf 
              - \frac{16}{9}\: \cf \ca \nf \nonumber \\[1.5mm]
  D_2^- & = &  \left[\, 81 + 14\,\z2 \right]\: \cf C_A^{\, 2}
              - \left[ \frac{152}{3} + 96\,\z2 \right] C_F^{\, 2} \ca
              - [ 60 - 104\, \z2 ]\: C_F^{\, 3} 
              - \frac{196}{9}\, \cf \ca \nf
 \nonumber \\ & & \mbox{}
             + \frac{80}{3}\, C_F^{\, 2} \nf 
             + \frac{4}{9} \cf n_{\! f}^{\, 2}  \nonumber \\[1.5mm]
  D_3^- & = &  \left[ \frac{3442}{27} + \frac{100}{3}\,\z2 
                      + 112\,\z3 \right] \cf C_A^{\, 2}
             + \left[ \frac{1850}{9} - \frac{680}{3}\,\z2 
                      - 336\,\z3 \right] C_F^{\, 2} \ca
             + \frac{88}{27}\, \cf n_{\! f}^{\, 2} \:\:\:\:
 \nonumber \\ & & \mbox{}
             - [ 286 - 192\,\z2 - 224\,\z3 ]\: C_F^{\, 3}
             + \left[ \frac{568}{9} + \frac{32}{3} \right] 
                 C_F^{\, 2} \nf 
             - \left[ \frac{2252}{27} - \frac{8}{3}\z2 \right] 
                \cf \ca \nf   \:\: ,
\eea
and
\bea
\label{eq:D-num}
  D_0^- &\cong & 1.43210  \nonumber \\
  D_1^- &\cong & 35.5556 - 3.16049\: \nf  \nonumber \\
  D_2^- &\cong & 399.205 - 39.7037\: \nf + 0.592592\: n_{\! f}^{\, 2}
  \nonumber \\
  D_3^- &\cong & 1465.93 - 172.693\: \nf + 4.345679\: n_{\! f}^{\, 2}
  \:\: .
\eea
The coefficients $D_0^\pm$ of the leading logarithms in 
Eqs.~(\ref{eq:D+exact}) and (\ref{eq:D-exact}) agree with the 
predictions in ref.~\cite{Blumlein:1996jp} based of the resummation of 
Ref.~\cite{Kirschner:1983di}.
Finally the small-$x$ expansion of $P^{\,(2)\rm s}_{\,\rm ns}(x)$ reads
\bea
\label{eq:DSexact}
  D_0^{\,\rm s} & = & \dabc2n\:\nf\:\: 
    \frac{1}{3}
  \nonumber \\[1mm]
  D_1^{\,\rm s} & = & \dabc2n\:\nf\: 
     \left( - \frac{2}{3} \right) 
  \nonumber \\[1mm]
  D_2^{\,\rm s} & = & \dabc2n\:\nf\: 
     ( 18 - 10\:\z2 )
  \nonumber \\[1mm]
  D_3^{\,\rm s} & = & \dabc2n\:\nf\:
     ( 56 + 2\:\z2 - 16\:\z3 ) \:\: ,
\eea
or, inserting the QCD value of $40/9$ for the group factor 
$d^{abc}d_{abc}/n_c$,
\bea
\label{eq:DSnum}
  D_0^{\, s} \: \cong \: + 1.48148\: \nf  \:\: & , & \quad
  D_1^{\, s} \: \cong \: - 2.96296\: \nf  \nonumber \\
  D_2^{\, s} \: \cong \: + 6.89182\: \nf  \:\: & , & \quad
  D_3^{\, s} \: \cong \: + 178.030\: \nf  \:\: .
\eea

The $n_{\! f}^{\: 0}$ and $n_{\! f}^{\, 1}$ parts of the functions
$P^{\,(2)\pm}_{\,\rm ns}(x)$ in Eqs.~(\ref{eq:Pqq2p}) and
(\ref{eq:Pqq2m}) are separately shown in Figs.~2 -- 4 together with
the approximate expressions derived in Ref.~\cite{vanNeerven:2000wp}
mainly from the \mbox{integer-$N$} results of
Refs.~\cite{Larin:1994vu,Larin:1997wd,Retey:2000nq}. Also shown for
the non-fermionic contributions in Figs.~2 and 3 are the successive
approximations by small-$x$ logarithms as defined in
Eq.~(\ref{eq:Plnx}) and the text below it.
As can be seen from Eqs.~(\ref{eq:D+num}) and (\ref{eq:D-num}), the
coefficients of $\ln^{\, k}x$ for $P^{\,(2)\pm}_{\,\rm ns}$ increase
sharply with decreasing power $k$. Consequently the shapes of the full
results in Figs.~2 and 3 are reproduced only after all logarithmically
enhanced terms have been included. Even then the \mbox{small-$x$}
approximations underestimate the complete results by factors as large as
2.7 and 2.0, respectively, for $P^{\,(2)+}_{\,\rm ns}$ and $P^{\,(2)-}
_{\,\rm ns}$ at $x = 10^{-4}$, rendering the small-$x$ expansion
(\ref{eq:Plnx}) ineffective for any practically relevant value of $x$.
Keeping only the Lx ($\,\ln^{\, 4}x$) or NLx ($\ln^{\, 4}x$ and
$\ln^{\, 3}x$) contributions leads to a reasonable description only
at extremely small values of $x$. Therefore, meaningful estimates 
of higher-order effects based on resumming leading (and subleading) 
logarithms in the small-$x$ limit appear to be difficult.

The new three-loop $n_{\!f}^{\,1}$ contribution $P^{\,(2)\rm s}
_{\,\rm ns}$ with the colour structure $d^{abc}d_{abc}/n_c$
is graphically displayed in Fig.~5 for $n_{\!f} = 1$. Rather
unexpectedly, also this function behaves like $\ln^{\, 4}x$ for
$x \ra 0$, and here the leading small-$x$ terms do indeed provide a
reasonable approximation. In fact, this function dominates the
small-$x$ behaviour of the non-singlet splitting functions, for
$n_{\!f}=4$ being, for example, about 7 times larger than
$P^{\,(2)\pm}_{\,\rm ns}(x)$ at $x = 10^{-4}$. The presence of a
(dominant) leading small-$x$ logarithm in a term unpredictable from
lower-order structures appears to call into question the very concept
of the small-$x$ resummation of the double logarithms
$\alpha_{\rm s}^{\, k+1}\,\ln^{\, 2k} x$.

\begin{figure}[p]
\label{pic:Pns2p0}
\vspace{-4mm}
\centerline{\epsfig{file=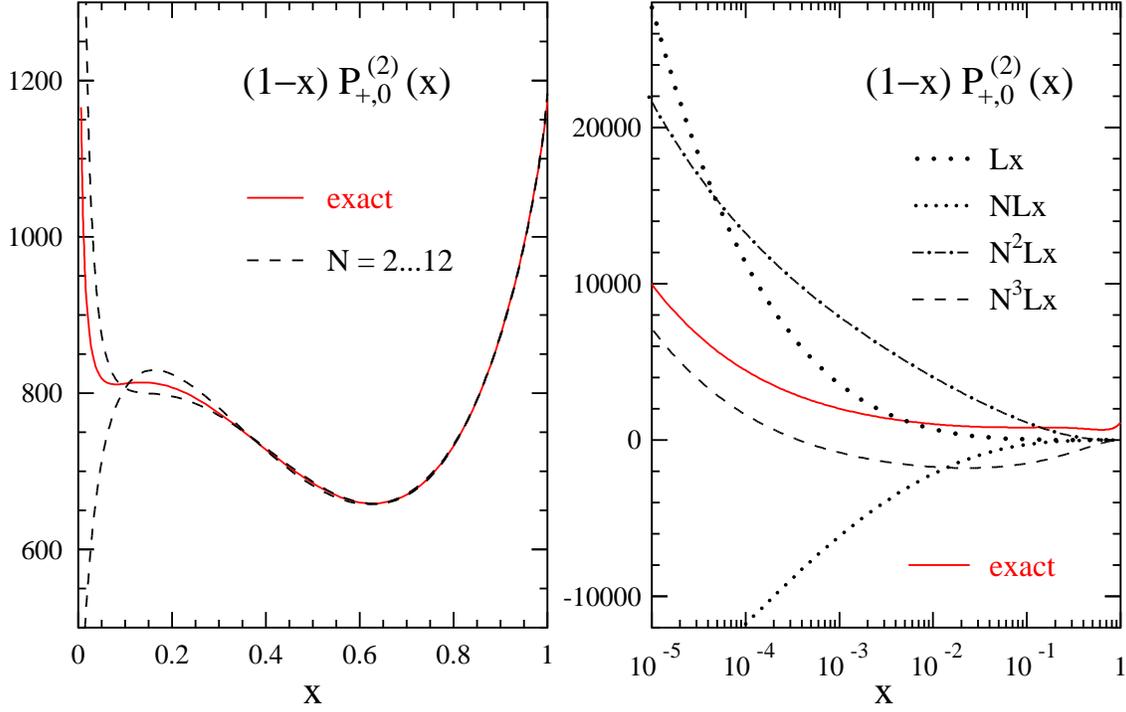,width=14.9cm,angle=0}}
\vspace{-2mm}
\caption{The $n_{\,\rm f\,}^{}$-independent three-loop contribution 
 $P_{+,0}^{\,(2)}(x)$ to the splitting function $P_{\rm ns}^{\, +}(x)$, 
 multiplied by $(1-x)$ for display purposes. Also shown in the left 
 part is the uncertainty band derived in Ref.~\cite{vanNeerven:2000wp} 
 from the lowest six even-integer moments
 \cite{Larin:1994vu,Larin:1997wd,Retey:2000nq}. In the right part our
 exact result is compared to the small-$x$ approximations defined in 
 Eq.~(\ref{eq:Plnx}) and the text below it.}
\vspace{1mm}
\end{figure}
\begin{figure}[p]
\label{pic:Pns2m0}
\vspace{-2mm}
\centerline{\epsfig{file=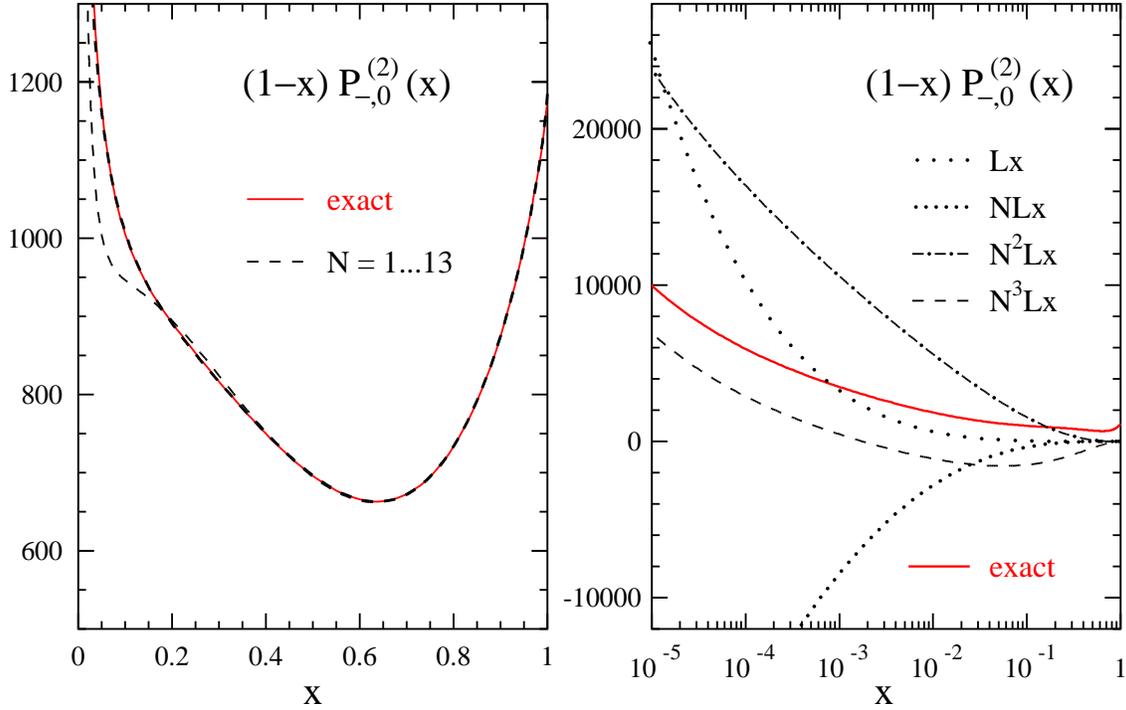,width=14.9cm,angle=0}}
\vspace{-2mm}
\caption{As Fig.~2, but for the splitting function 
 $P_{\rm ns}^{\,-}(x)$. The first seven odd moments underlying the 
 previous approximations \cite{vanNeerven:2000wp} also shown in the 
 left part have been computed in Ref.~\cite{Retey:2000nq}.}
\vspace{1mm}
\end{figure}

\begin{figure}[p]
\label{pic:Pns2nf}
\vspace{-2mm}
\centerline{\epsfig{file=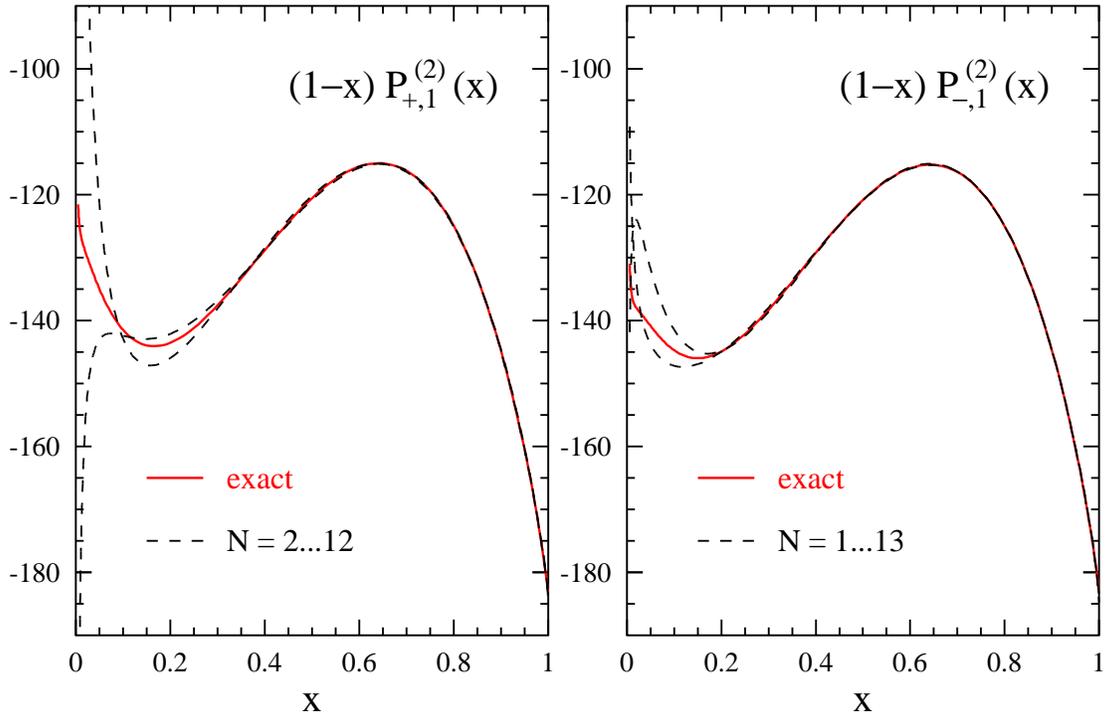,width=15.0cm,angle=0}}
\vspace{-2mm}
\caption{The $n_{\,\rm f\,}^{\, 1}$ three-loop contributions 
 $P_{\pm,1}^{\,(2)}(x)$ to the splitting functions 
 $P_{\rm ns}^{\,\pm}(x)$,  compared to the uncertainty bands of 
 Ref.~\cite{vanNeerven:2000wp} based on the integer moments calculated
 in Refs.\ \cite{Larin:1994vu,Larin:1997wd,Retey:2000nq}.} 
\vspace{1mm}
\end{figure}
\begin{figure}[p]
\label{pic:Pns2s}
\vspace{-2mm}
\centerline{\epsfig{file=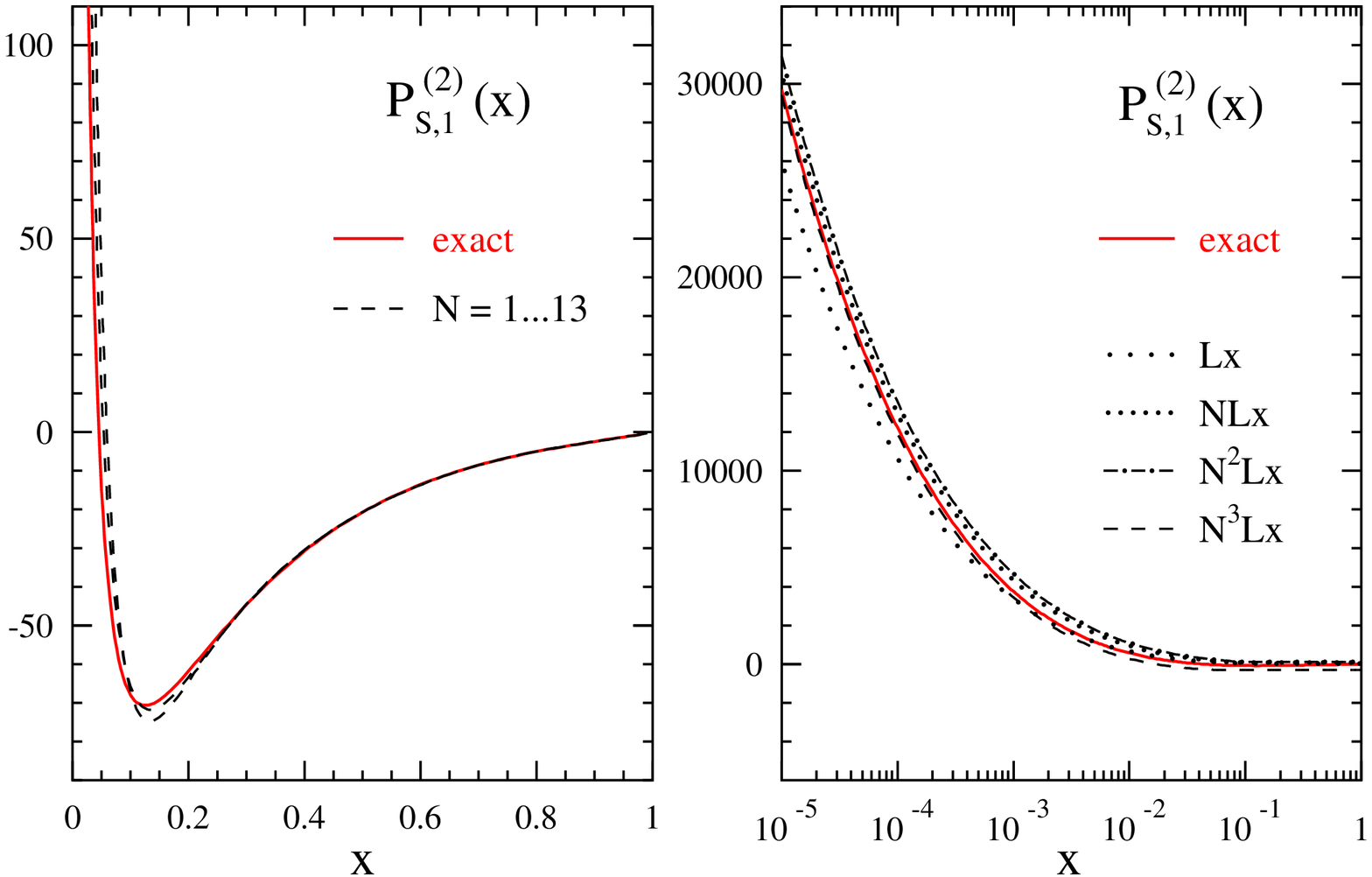,width=15.0cm,angle=0}}
\vspace{-2mm}
\caption{The first non-vanishing contribution $P_{{\rm s},1}^{\,(2)}
 (x)$ to the splitting functions $P_{\rm ns}^{\: s}(x)$, compared to
 the approximations of Ref.~\cite{vanNeerven:2000wp} (where, assuming
 the completeness of the resummation 
 \cite{Kirschner:1983di,Blumlein:1996jp}, the possibility of a 
 $\ln^{\,4}x$ term was disregarded) and to the small-$x$ expansion in 
 powers of $\ln x$.}
\vspace{1mm}
\end{figure}

In view of the length and complexity of the exact expressions for
the functions $P^{\, (2)i}_{\rm ns}(x)$, it is useful to have 
at ones disposal also compact approximate representations involving, 
besides powers of $x$, only simple functions like the $+$-distribution 
and the end-point logarithms
\beq
\label{eq:logs}
  \DD_{\,0} \: = \: 1/(1-x)_+ \: ,
  \quad L_1 \: = \: \ln (1-x) \: ,
  \quad L_0 \: = \: \ln x \:\: .
\eeq
Inserting the numerical values of the QCD colour factors, 
$P^{\,(2)+}_{\,\rm ns}$ in Eq.~(\ref{eq:Pqq2p}) can be represented~by 
\bea
\label{eq:P+appr}
  P^{(2)+}_{\,\rm ns}(x)\!\!\! & \cong & \mbox{} \!\!
     + 1174.898\: \DD_0 + 1295.384\: \delta (1-x) + 714.1\: L_1
     + 1641.1 - 3135\: x + 243.6\: x^2 
  \nonumber \\[1mm] & & \mbox{} \!\!
     - 522.1\: x^3 + L_0 L_1 [563.9 + 256.8\: L_0] + 1258\: L_0 
     + 294.9\: L_0^2 + 800/27\: L_0^3 + 128/81\: L_0^4
  \nonumber \\ &+& \mbox{} \nf \:\Big(
     - 183.187\: \DD_0 - 173.927\: \delta (1-x)
     - 5120/81\: L_1 - 197.0 + 381.1\: x + 72.94\: x^2
  \nonumber \\[-0.5mm] & & \mbox{} \quad 
     + 44.79\: x^3 - 1.497\: xL_0^3 - 56.66\: L_0 L_1 
     - 152.6\: L_0 - 2608/81\: L_0^2 - 64/27\: L_0^3\, \Big) 
  \nonumber \\ &+& \mbox{} \n2f \:\Big(
     - \DD_0 - (51/16 + 3\,\z3 - 5\,\z2) \: \delta (1-x) 
     + x\,(1-x)^{-1} L_0\, (3/2\: L_0 + 5) + 1
  \nonumber \\[-0.5mm] & & \mbox{} \quad
     + (1-x)\, (6 + 11/2\: L_0 + 3/4\: L_0^2)\, \Big) \:\, 64/81 
  \:\: .
\eea
A corresponding parametrization of $P^{\,(2)-}_{\,\rm ns}$ in 
Eq.~(\ref{eq:Pqq2m}) is given by
\bea
\label{eq:P-appr}
  P^{(2)-}_{\,\rm ns}(x)\!\!\! & \cong & \mbox{} \!\!
     + 1174.898\: \DD_0 + 1295.470\: \delta (1-x) + 714.1\: L_1
     + 1860.2 - 3505\: x + 297.0\: x^2 
  \nonumber \\[1mm] & & \mbox{} \!\!
     - 433.2\: x^3 + L_0 L_1 [684 + 251.2\: L_0 ]+ 1465.2\: L_0 
     + 399.2\: L_0^2 + 320/9\: L_0^3 + 116/81\: L_0^4
  \nonumber \\ &+& \mbox{} \nf \:\Big(
     - 183.187\: \DD_0 - 173.933\: \delta (1-x)
     - 5120/81\: L_1 - 216.62 + 406.5\: x + 77.89\: x^2
  \nonumber \\[-0.5mm] & & \mbox{} \quad 
     + 34.76\: x^3 - 1.136\: xL_0^3 - 65.43\: L_0 L_1 
     - 172.69\: L_0 - 3216/81\: L_0^2 - 256/81\: L_0^3\, \Big) 
  \nonumber \\ &+& \mbox{} \n2f \:\Big(
     - \DD_0 - (51/16 + 3\,\z3 - 5\,\z2) \: \delta (1-x) 
     + x\,(1-x)^{-1} L_0\, (3/2\: L_0 + 5) + 1
  \nonumber \\[-0.5mm] & & \mbox{} \quad
     + (1-x)\, (6 + 11/2\: L_0 + 3/4\: L_0^2)\, \Big) \:\, 64/81 
  \:\: .
\eea
Finally the splitting function $P^{\,(2){\rm s}}_{\,\rm ns}$ in 
Eq.~(\ref{eq:Pqq2s}) can be approximated by
\bea
\label{eq:Psappr}
  P^{(2)\rm s}_{\,\rm ns}(x)\!\! & \cong & \nf \:\Big(
     [\, L_1 ( - 163.9\: x^{-1} - 7.208\: x) + 151.49 +  44.51\: x
     - 43.12\: x^2 + 4.82\: x^3\, ]\, [1-x\, ]  \quad
  \nonumber \\[-0.5mm] & & \mbox{} \!\!\!
     + L_0 L_1 [ - 173.1 + 46.18\: L_0 ] + 178.04\: L_0
     + 6.892\: L_0^2 + 40/27\: [ L_0^4 - 2\: L_0^3\, ] \, \Big) 
  \:\ . \quad\quad
\eea
The identical $\n2f$ parts of $P^{(2)\pm}_{\,\rm ns}$, the 
$+$-distribution contributions (up to a numerical truncation of the 
coefficients involving $\zeta_{i\,}$), and the rational coefficients of 
the (sub-)leading regular end-point terms are exact in 
Eqs.~(\ref{eq:P+appr}) -- (\ref{eq:Psappr}). The remaining coefficients 
have been determined by fits to the exact results, for which we have 
used the {\sc Fortran} package of Ref.~\cite{Gehrmann:2001pz}. Except for
$x$ values very close to zeros of $P^{(2)i}_{\rm ns}(x)$, the above 
parametrizations deviate from the exact expressions by less than one 
part in thousand, which should be sufficiently accurate for foreseeable 
numerical applications. For a maximal accuracy for the convolutions 
with the quark densities, also the coefficients of $\delta (1-x)$ have 
been slightly adjusted, by 0.02\% or less, using low integer moments.
Also the complex-$N$ moments of the splitting functions 
can be readily obtained to a perfectly sufficient
accuracy using Eqs.~(\ref{eq:P+appr}) -- (\ref{eq:Psappr}). The
Mellin transform of these parametrizations involve only simple
harmonic sums $S_{m>0}(N)$ (see, e.g, the appendix of Ref.~\cite
{Blumlein:1998if}) of which the analytic continuations in terms
of logaritmic derivatives of Euler's $\Gamma$-function are well known.
%
%
\setcounter{equation}{0}
\section{Numerical implications}
\label{sec:sresults}
%
%
In this section we illustrate the effect of our new three-loop 
splitting functions $P^{(2)\pm,\rm v}_{\,\rm ns}(x)$ on the evolution 
(\ref{eq:evol}) of the non-singlet combinations $q_{\,\rm ns}^{\,\pm,
\rm v}(x,\mu_f^{\,2})$ of the quark and antiquark distributions. 
For all figures we employ the same schematic, but characteristic model 
distribution,
\beq
\label{eq:shape}
  xq_{\,\rm ns}^{\,\pm,\rm v}(x,\mu_{0}^{\,2}) \: = \: 
  x^{\, a} (1-x)^b \:\: 
\eeq
with
\beq
 a \: = \: 0.5 \:\:, \quad b \: = \: 3 \:\: , 
\eeq
facilitating a direct comparison of the various splitting functions
contributing to Eq.~(\ref{eq:evol}). For the same reason the reference
scale is specified by an order-independent value for the strong 
coupling constant usually chosen as
\beq
  \as (\mu_{0}^{\,2}) \: = \: 0.2 \:\: .
\eeq
This value corresponds to $\mu_{0}^{\,2} \,\simeq\, 25\ldots 50$ 
GeV$^2$ for $\as (M_Z^{\, 2}) = 0.114 \ldots 0.120$ beyond the leading
order, a scale region relevant for deep-inelastic scattering both
at fixed-target experiments and, for much smaller $x$, at the {\it ep}
collider HERA. Our default for the number of effectively massless
flavours is $\nf =4$. The normalization of $q_{\,\rm ns}^{\: i}$
is irrelevant for our purposes, as we consider only the logaritmic
derivatives $\dot{q}_{\rm ns}^{\: i} \equiv d \ln q_{\rm ns}^{\: i}/ 
d\ln \mu_f^{\,2}$. 

\begin{figure}[hbt]
\label{pic:dq+exp}
\vspace{-2mm}
\centerline{\epsfig{file=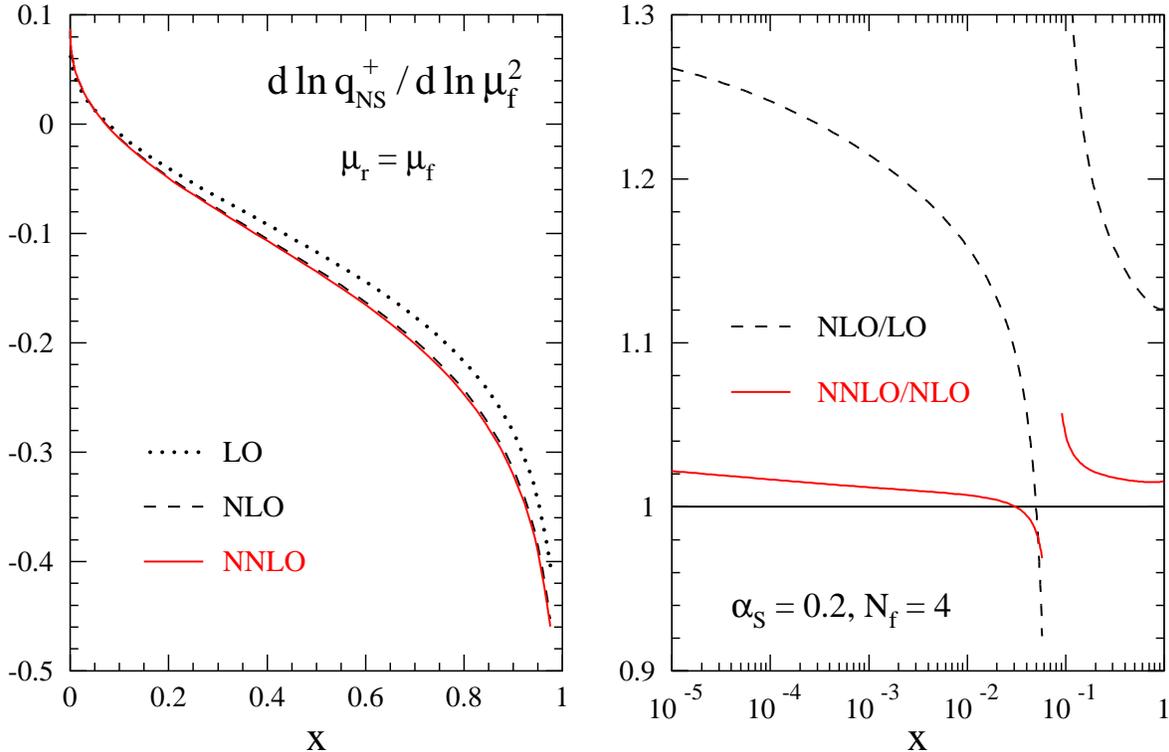,width=16cm,angle=0}}
\vspace{-2mm}
\caption{The perturbative expansion of the logarithmic scale derivative 
 $\,d \ln q_{\,\rm ns}^{\, +}/ d\ln \mu_f^{\,2}\,$ for a characteristic
 non-singlet quark distribution $xq_{\rm ns}^{\, +}\, =\, x^{\, 0.5} 
 (1-x)^3$ at the standard scale $\mu_r = \mu_f$.}
\end{figure}

\begin{figure}[tbh]
\label{pic:dq-exp}
\vspace*{-2mm}
\centerline{\epsfig{file=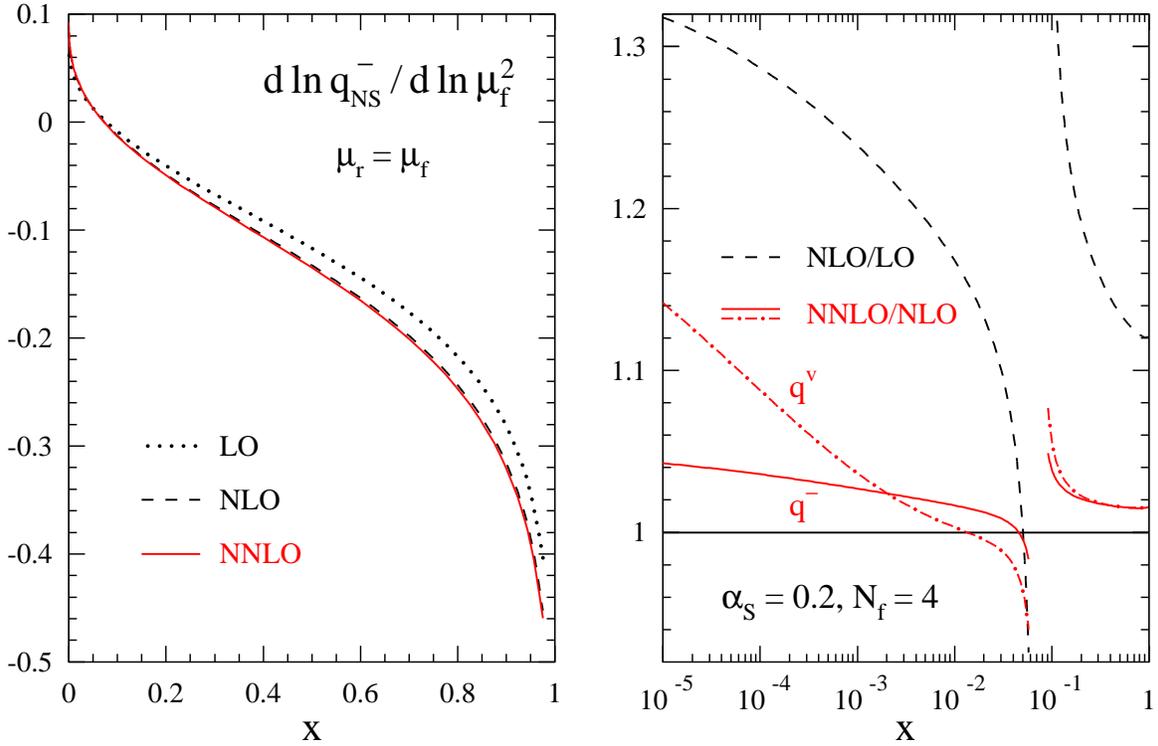,width=15.8cm,angle=0}}
\vspace{-2mm}
\caption{As Fig.~6, but for the scale derivatives of the two other
 non-singlet combinations $q_{\,\rm ns}^{\, -,{\rm v}}$.}
\vspace{-2mm}
\end{figure}

The scale derivatives of the three non-singlet distributions are
graphically displayed in Figs.~6 and 7 over a wide region of $x$. At
large $x$ the NNLO corrections are very similar in all cases, amounting 
to 2\% or less for $x\geq 0.2$, thus being smaller than the NLO 
corrections by a factor of about eight. The same suppression factor is 
also found for $q_{\,\rm ns}^{\,-}(x)$ in the region $10^{-5} \lsim x
\lsim 10^{-2}$. The NNLO effects are even smaller for $q_{\,\rm ns}^{\, 
+}$ at small $x$, but considerably larger for $q_{\,\rm ns}^{\,\rm v}$ 
at $x< 10^{-3}$. For example, at $x\simeq 10^{-4}$, where $P^{(2)\rm v}
_{\rm ns}(x)$ exceeds $P^{(2)-}_{\rm ns}(x)$ by a factor of about 8 as 
discussed in the paragraph above Eq.~(\ref{eq:logs}), the ratio of the 
corresponding corrections in Fig.~7 amounts to 2.5. Recall that the 
scale derivatives (\ref{eq:evol}) do not probe the splitting functions 
locally in $x$ due to the presence of the Mellin convolution.

The numerical values for $\dot{q}_{\rm ns}^{\:\rm v}(x,\mu_{0}^{\,2})$
are presented in Tab.~1 for four characteristic values of $x$. Also
illustrated in this table is the dependence of the results on the shape 
of the initial distribution, the number of flavours and the value of 
the strong coupling constant. The relative corrections are rather
weakly dependent of the large-$x$ power $b$ in Eq.~(\ref{eq:shape}).
They increase at small $x$ with increasing small-$x$ power $a$, i.e., 
with decreasing size of $q_{\rm ns}^{\:\rm v}$. At large $x$, where
the $\nf\, d^{abc\,}d_{abc}/n_c$ contribution $P^{\,\rm s}
_{\,\rm ns}$ is negligible, the NNLO corrections decrease with 
increasing $\nf$. At small-$x$ this decrease is overcompensated in 
$\dot{q}_{\rm ns}^{\:\rm v}$ by the effect of $P^{\,\rm s}_{\rm ns}$. 
Except for very small momentum fractions $x \lsim 10^{-3}$ (where the
non-singlet quark densities play a minor role for most important 
observables) the NNLO corrections amount to 15\% or less even for a 
strong coupling constant as large as \mbox{$\as = 0.5$}. Hence the 
non-singlet evolution at intermediate and large $x$ appears to remain 
perturbative down to very low scales as used in the phenomenological
analyses of Refs.~\cite{Gluck:1995uf,Gluck:1998xa} and in non-perturbative 
studies of the initial distributions like those of Refs.\
\cite{Weigel:1997kw,Diakonov:1996sr,Diakonov:1997vc,Pobylitsa:1998tk,%
Schroeder:1999fr,Weigel:1999pc}.

\begin{table}[htp]
\label{table1}
\begin{center}
\begin{tabular}{||c||r|r|r||r|r||c||}
\hline \hline
 & & & & & & \\[-0.3cm]
\multicolumn{1}{||c||}{$x$} &
\multicolumn{1}{c|} {LO} &
\multicolumn{1}{c|} {NLO} &
\multicolumn{1}{c||} {NNLO} &
\multicolumn{1}{c|} {$r_1$} &
\multicolumn{1}{c||}{$r_{\,2}$} &
\multicolumn{1}{c||}{$r_{\,2}/r_1$} \\[0.5mm] \hline \hline
\multicolumn{7}{||c||}{} \\[-3mm]
\multicolumn{7}{||c||}{ default (Fig.~7) } \\
\multicolumn{7}{||c||}{} \\[-0.3cm] \hline \hline
 & & & & & & \\[-0.2cm]
$ 10^{-4}$ &
$ 6.546\cdot 10^{-2}$ &$ 8.424\cdot 10^{-2}$ &$ 9.163\cdot 10^{-2}$ &
  0.287 & 0.088 & 0.31 \\
  0.002  &
$ 5.632\cdot 10^{-2}$ &$ 6.875\cdot 10^{-2}$ &$ 7.041\cdot 10^{-2}$ &
  0.221 & 0.024 & 0.11 \\
  0.25  &
$-5.402\cdot 10^{-2}$ &$-6.331\cdot 10^{-2}$ &$-6.457\cdot 10^{-2}$ &
  0.172 & 0.020 & 0.12 \\
  0.75  &
$-1.949\cdot 10^{-1}$ &$-2.189\cdot 10^{-1}$ &$-2.222\cdot 10^{-1}$ &
  0.123 & 0.015 & 0.12 \\[1mm]
\hline \hline
\multicolumn{7}{||c||}{} \\[-3mm]
\multicolumn{7}{||c||}{ $a = 0.8$ } \\
\multicolumn{7}{||c||}{} \\[-0.3cm] \hline \hline
 & & & & & & \\[-0.2cm]
$ 10^{-4}$ &
$ 1.660\cdot 10^{-1}$ &$ 2.351\cdot 10^{-1}$ &$ 2.818\cdot 10^{-1}$ &
  0.417 & 0.198 & 0.48 \\
 0.002   &
$ 1.249\cdot 10^{-1}$ &$ 1.583\cdot 10^{-1}$ &$ 1.650\cdot 10^{-1}$ &
  0.268 & 0.042 & 0.16 \\
  0.25  &
$-4.352\cdot 10^{-2}$ &$-5.171\cdot 10^{-2}$ &$-5.283\cdot 10^{-2}$ &
  0.188 & 0.022 & 0.12 \\
  0.75  &
$-1.930\cdot 10^{-1}$ &$-2.168\cdot 10^{-1}$ &$-2.200\cdot 10^{-1}$ &
  0.123 & 0.015 & 0.12 \\[1mm]
\hline \hline
\multicolumn{7}{||c||}{} \\[-3mm]
\multicolumn{7}{||c||}{ $b = 5$ } \\
\multicolumn{7}{||c||}{} \\[-0.3cm] \hline \hline
 & & & & & & \\[-0.2cm]
$ 10^{-4}$ &
$ 6.474\cdot 10^{-2}$ &$ 8.278\cdot 10^{-2}$ &$ 8.917\cdot 10^{-2}$ &
  0.279 & 0.077 & 0.28 \\
  0.002  &
$ 5.324\cdot 10^{-2}$ &$ 6.432\cdot 10^{-2}$ &$ 6.546\cdot 10^{-2}$ &
  0.208 & 0.018 & 0.09 \\
  0.25  &
$-7.835\cdot 10^{-2}$ &$-9.022\cdot 10^{-2}$ &$-9.180\cdot 10^{-2}$ &
  0.151 & 0.018 & 0.12 \\
  0.75  &
$-2.300\cdot 10^{-1}$ &$-2.580\cdot 10^{-1}$ &$-2.619\cdot 10^{-1}$ &
  0.122 & 0.015 & 0.12 \\[1mm]
\hline \hline
\multicolumn{7}{||c||}{} \\[-3mm]
\multicolumn{7}{||c||}{ $\nf = 3$ } \\
\multicolumn{7}{||c||}{} \\[-0.3cm] \hline \hline
 & & & & & & \\[-0.2cm]
$ 10^{-4}$ &
$ 6.546\cdot 10^{-2}$ &$ 8.480\cdot 10^{-2}$ &$ 9.187\cdot 10^{-2}$ &
  0.295 & 0.083 & 0.28 \\
  0.002  &
$ 5.632\cdot 10^{-2}$ &$ 6.942\cdot 10^{-2}$ &$ 7.174\cdot 10^{-2}$ &
  0.233 & 0.033 & 0.14 \\
  0.25  &
$-5.402\cdot 10^{-2}$ &$-6.406\cdot 10^{-2}$ &$-6.588\cdot 10^{-2}$ &
  0.186 & 0.028 & 0.15 \\
  0.75  &
$-1.949\cdot 10^{-1}$ &$-2.219\cdot 10^{-1}$ &$-2.269\cdot 10^{-1}$ &
  0.139 & 0.023 & 0.16 \\[1mm]
\hline \hline
\multicolumn{7}{||c||}{} \\[-3mm]
\multicolumn{7}{||c||}{ $\nf = 3\:$ and $\:\as = 0.5$ } \\
\multicolumn{7}{||c||}{} \\[-0.3cm] \hline \hline
 & & & & & & \\[-0.2cm]
$ 10^{-4}$ &
$ 1.636\cdot 10^{-1}$ &$ 2.845\cdot 10^{-1}$ &$ 3.949\cdot 10^{-1}$ &
  0.739 & 0.388 & 0.53 \\
  0.002  &
$ 1.408\cdot 10^{-1}$ &$ 2.227\cdot 10^{-1}$ &$ 2.589\cdot 10^{-1}$ &
  0.581 & 0.163 & 0.28 \\
$ 0.25   $ &
$-1.350\cdot 10^{-1}$ &$-1.978\cdot 10^{-1}$ &$-2.262\cdot 10^{-1}$ &
  0.465 & 0.144 & 0.31 \\
$ 0.75   $ &
$-4.871\cdot 10^{-1}$ &$-6.563\cdot 10^{-1}$ &$-7.346\cdot 10^{-1}$ &
  0.347 & 0.119 & 0.34 \\[1mm]
\hline \hline
\end{tabular}
\end{center}
\caption{The LO, NLO and NNLO logarithmic derivatives $\dot{q}_{\rm ns}
 ^{\,\rm v }\equiv d \ln q_{\rm ns}^{\: v}/d\ln \mu_f^{\,2}$ at four 
 representative values of $x$, together with the ratios $r_n = 
 {\rm N^nLO} / {\rm N^{n-1}LO}- 1\,$ for the default input parameters
 specified in the first paragraph of this section and some variations 
thereof.}
\end{table}

Another conventional way to assess the reliability of perturbative 
calculations is to investigate the stability of the results under 
variations of the renormalization scale $\mu_r$. For $\mu_r \neq \mu_f$
the expansion in Eq.~(\ref{eq:evol}) has to be replaced by
\bea
  P_{\rm ns}^{\, i}(\mu_f,\mu_r)
  &\! =\! & \quad
      a_s(\mu_r^2) \, P_{\rm ns}^{(0)}  \:\: + \:\: 
      a_s^2(\mu_r^2) \, \left( P_{\rm ns}^{(1),i} - \beta_0
      P_{\rm ns}^{(0)} \ln \frac{\mu_f^2}{\mu_r^2} \right) \:  \\
  & & \mbox{}\!\!\!
    + \:  a_s^3(\mu_r^2) \, \left( P_{\rm ns}^{(2),i}
    - \bigg\{ \beta_1 P_{\rm ns}^{(0)} + 2\beta_0 P_{\rm ns}^{(1),i}
   \bigg\} \ln \frac{\mu_f^2}{\mu_r^2} 
    + \beta_0^2 P_{\rm ns}^{(0)} \ln^2 \frac{\mu_f^2}{\mu_r^2}\,
      \right) + \ldots \nn \:\: ,
\eea
where $\beta_k$ represent the \MSb\ expansion coefficients of the 
$\beta$-function of QCD \
\cite{Caswell:1974gg,Jones:1974mm,Tarasov:1980au,Larin:1993tp}.

In Fig.~8 the consequences of varying $\mu_r$ over the rather wide 
range $\frac{1}{8}\,\mu_f^2 \,\leq\, \mu_r^2 \,\leq\, 8 \mu_f^2$ are 
displayed for $\dot{q}_{\rm ns}^{\, +}$ at six representative values of 
$x$. The scale dependence is considerably reduced by including the 
third-order corrections over the full $x$-range. At NNLO both the
points of fastest apparent convergence and the points of minimal
$\mu_r$-sensitivity, $\partial\dot{q}_{\rm ns}^{\,+}/\partial\mu_r =0$, 
are rather close to the `natural' choice $\mu_r = \mu_f$ for the 
renormalization scale.

\begin{figure}[tbh]
\label{pic:scales1}
\vspace{-2mm}
\centerline{\epsfig{file=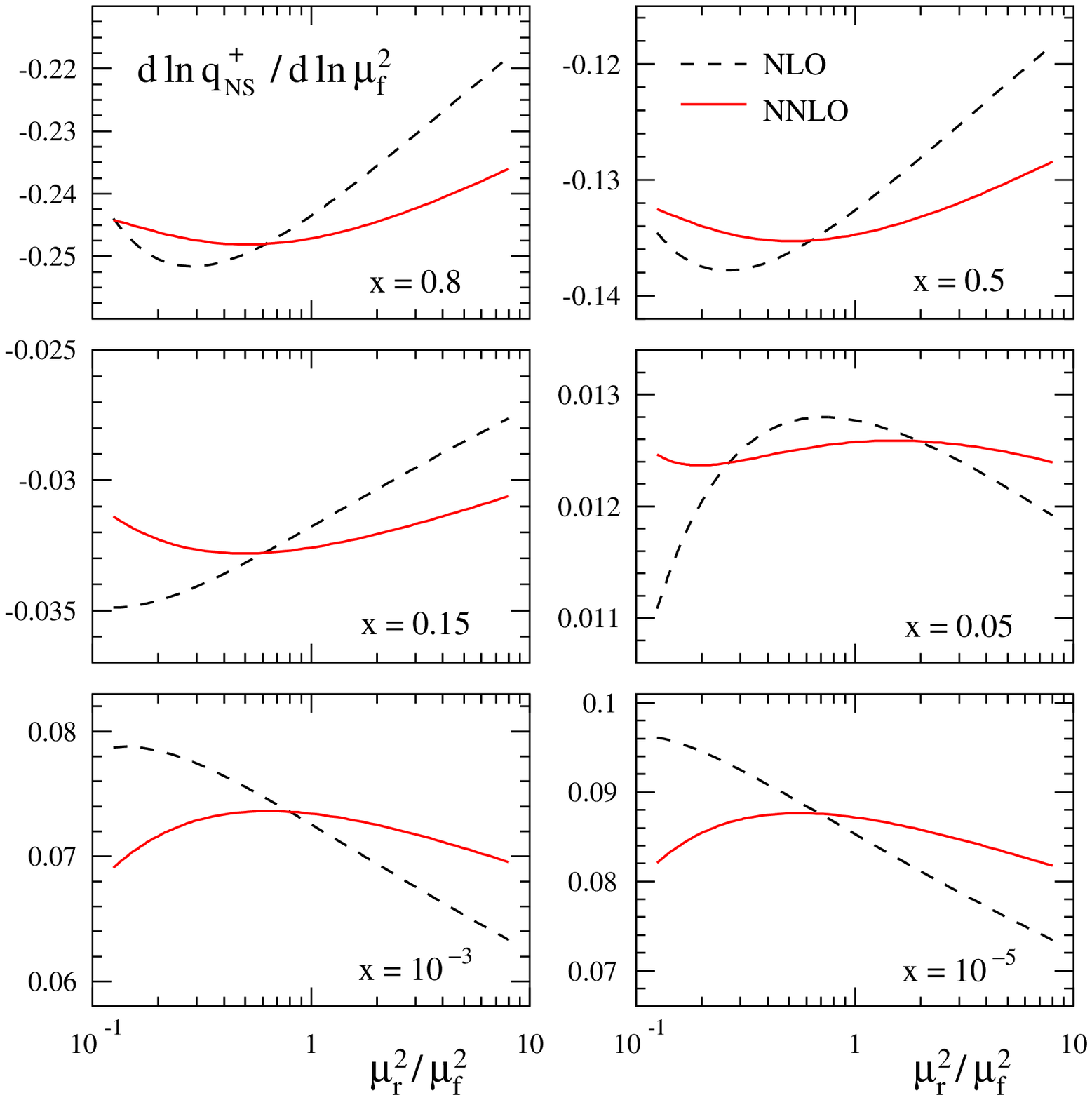,width=14.4cm,angle=0}}
\vspace{-2mm}
\caption{The dependence of the NLO and NNLO predictions for $\dot{q}_
 {\rm ns}^{\, +} \:\equiv\: d \ln q_{\rm ns}^{\, +}/ d\ln \mu_f^2$ on
 the renormalization scale $\mu_r$ for six typical values of $x$.
 The initial conditions are as in Fig.~\ref{pic:dq+exp}.}
\end{figure}

The relative scale uncertainties of the average results, conventionally
estimated by
\beq
\label{eq:screl}
 \Delta \dot{q}_{\rm ns}^{\, i} \: \equiv \:
 \frac{\max\, [ \dot{q}_{\rm ns}^{\, i}(x,\mu_r^2 = \frac{1}{4}
 \mu_f^2 \ldots 4\mu_f^2)] - \min\, [\dot{q}_{\rm ns}^{\, i} (x,
 \mu_r^2 = \frac{1}{4}\mu_f^2 \ldots 4 \mu_f^2)] }
 { 2\, |\, {\rm average}\, [\dot{q}_{\rm ns}^{\, i}(x, \mu_r^2 =
 \frac{1}{4}\mu_f^2 \ldots 4 \mu_f^2)]\, | }
\eeq
is shown in Fig.~9 for all three cases $i = \pm,$v. These uncertainty
estimates amount to 2\% or less except for $x \lsim 10^{-3}$, an 
improvement by more than a factor of three with respect to the 
corresponding NLO results. Taking into account also the apparent
convergence of the series in Figs.~6 and 7, it is not unreasonable 
to expect that the effect of the four-loop non-singlet splitting functions 
--- which most likely will remain uncalculated for quite some time --- 
will be less than 1\% for $x > 10^{-3}$. 
This expectation is consistent with the Pad\'{e} estimates
of $P_{\rm ns}^{\,(3)i}$ employed in Ref.~\cite{vanNeerven:2001pe} for 
the N$^3$LO large-$x$ evolution of the deep-inelastic structure 
functions $F_2$ and $F_3$. At very small values of $x$ the higher-order 
corrections will presumably be considerably larger.

\begin{figure}[tbh]
\label{pic:scales2}
\vspace{-2mm}
\centerline{\epsfig{file=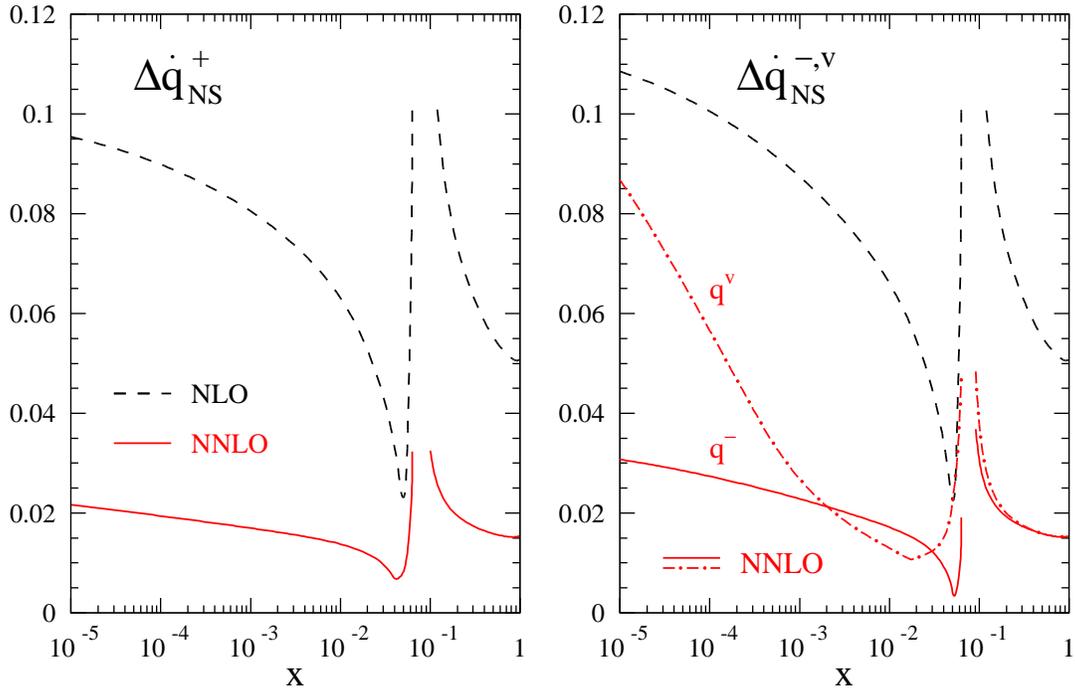,width=14.6cm,angle=0}}
\vspace{-2mm}
\caption{The renormalization scale uncertainty of the NLO and NNLO 
 predictions for the scale derivative of $q_{\rm ns}^{\,i}$, $i=\pm,V$,
 as obtained from the quantity $\Delta \dot{q}_{\rm ns}^{\, i}$ defined 
 in Eq.~(\ref{eq:screl}). Here and in Figs.~6 and 7 the spikes close
 to $x=0.1$ reflect the sign-change of $\dot{q}_{\rm ns}^{\, i}$ and
 do not constitute appreciable absolute corrections and uncertainties.}
\end{figure}
%
%
\section{Summary}
\label{sec:summary}
%
%
We have calculated the complete third-order contributions to the
splitting functions governing the evolution of unpolarized non-singlet
parton distribution in perturbative QCD. Our calculation is performed
in Mellin-$N$ space and follows the previous fixed-$N$ computations
\cite{Larin:1994vu,Larin:1997wd,Retey:2000nq} inasmuch as we compute the
partonic structure functions in deep-inelastic scattering at even or
odd $N$ using the optical theorem and a dispersion relation as
discussed in \cite{Larin:1997wd}. Our calculation, however, is not 
restricted to low fixed values of $N$ but provides the complete 
$N$-dependence from which the \mbox{$x$-space} splitting functions can 
be obtained by a (by now) standard Mellin inversion. 
This progress has been made possible by an improved understanding of 
the mathematics of harmonic sums, difference equations and harmonic 
polylogarithms \cite{Vermaseren:1998uu,Remiddi:1999ew,Moch:1999eb}, and 
the implementation of corresponding tools, together with other new 
features \cite{Vermaseren:2002rp}, in the symbolic manipulation program 
{\sc Form}~\cite{Vermaseren:2000nd} which we have employed to handle 
the almost prohibitively large intermediate expressions.

Our results have been presented in both Mellin-$N$ and Bjorken-$x$
space, in the latter case we have also provided easy-to-use accurate
parametrizations. Our results agree with all partial results available 
in the literature, in particular we reproduce the lowest seven even- or 
odd-integer moments computed before \cite{Larin:1994vu,Larin:1997wd,%
Retey:2000nq}. We also agree with the resummation predictions
\cite{Kirschner:1983di,Blumlein:1996jp} for the leading small-$x$ 
logarithms $\ln^{\,4}x$ of the splitting functions $P_{\rm ns}^{\,+}
(x)$ and $P_{\rm ns}^{\,-}(x)$ governing the evolution of flavour 
differences of quark-antiquark sums and differences. 
However, an unpredicted term of the same size is found also for the new 
$d^{abc\,}d_{abc}/n_c$ contributions $P_{\rm ns}^{\,\rm s}$ 
to the splitting function for the total valence distribution. At large
$x$ we find that the coefficients of the leading integrable term
$\ln (1-x)$ at order $n$ is proportional to the coefficient of the
(only) $+$-distribution $1/(1-x)_+$ at order $n-1$, a result that
seems to point to a yet unexplored structure.

We have investigated the numerical impact of the three-loop (NNLO)
contributions on the evolution of the various non-singlet densities.
The effect of the new contribution $P_{\rm ns}^{\,-}(x)$ is very small 
at large $x$ but rises sharply towards $x\ra 0$, reaching 10\% for a 
standard Regge-inspired $\sqrt{x}$ initial distributions at 
$x \simeq 10^{-5}$.
At $x > 10^{-3}$, on the other hand, the perturbative expansion for the 
scale dependences $d \ln q_{\rm ns}(x,\mu_f^{\,2})/ d\ln \mu_f^{\,2}$ 
appear to be very well convergent. 
For $\as = 0.2$, for example, the NNLO corrections amount to 2\% or 
less for four flavours, a factor of about 8 less than the NLO 
contributions. Also the variation of the renormalization scale leads 
to effects of about $\pm 2\%$ at NNLO in this region of $x$. 
Corrections of this size are comparable to the dependence of the 
predictions on the number of quark flavours, rendering a proper 
treatment of charm effects rather important even for large-$x$ 
non-singlet quantities, see Refs.~\cite{Laenen:1993zk,Chuvakin:1999nx} 
and references therein.
 
{\sc Form} files of our results, and {\sc Fortran} subroutines of our
exact and approximate $x$-space splitting functions can be obtained 
from the preprint server \ {\tt http://arXiv.org} by downloading the source.
Furthermore they are available from the authors upon request.
\subsection*{Acknowledgments}
The preparations for this calculation have been started by J.V.
following a suggestion by S.A.~Larin. For stimulating discussions
during various stages of this project we would like to thank, in
chronological order, S.A.~Larin, F.~J.~Yndurain, E.~Remiddi, 
E.~Laenen, W.~L.~van Neerven, P.~Uwer, S.~Weinzierl and J.~Bl\"umlein. 
M.~Zhou has contributed some {\sc Form} routines during an early stage 
of the calculation.
The work of S.M. has been supported in part by Deutsche 
Forschungsgemeinschaft in Sonderforschungsbereich/Transregio 9.
The work of J.V. and A.V. has been part of the research program of the
Dutch Foundation for Fundamental Research of Matter (FOM).
 

\end{document}